\documentclass[twocolumns,a4paper]{aa}
\usepackage{graphicx}
\usepackage{txfonts}
\usepackage{color}
\usepackage{natbib}
\usepackage{hyperref}
\usepackage{placeins}

\usepackage{amsmath}
\usepackage{float}
\usepackage{xcolor}
\usepackage{ulem}

\definecolor{blue}{RGB}{0,0,255}
\definecolor{red}{RGB}{255,0,0}
\definecolor{green}{RGB}{0,255,0}
\definecolor{limegreen}{RGB}{50,205,50}

\begin{document}

\title{The role of the Lorentz force in sunspot equilibrium}
\author{J.M.~Borrero\inst{1} \and A. Pastor Yabar\inst{2} \and M. Schmassmann\inst{1} \and
M. Rempel\inst{3} \and M. van Noort\inst{4} \and M. Collados\inst{5,6}}
\institute{Institut f\"ur Sonnenphysik, Sch\"oneckstr. 6, D-79104, Freiburg, Germany
\and
Institute for Solar Physics, Department of Astronomy, Stockholm University, AlbaNova University 
Centre, 10691 Stockholm, Sweden
\and 
High Altitude Observatory, NSF National Center for Atmospheric Research, 3080 Center Green Dr., Boulder 80301, USA
\and
Max Planck Institute for Solar System Research, Justus-von-Liebig 3, D-37077, G\"ottingen, Germany
\and
Instituto de Astrof{\'\i}sica de Canarias, Avd. V{\'\i}a L\'actea s/n, E-38205, La Laguna, Spain
\and
Departamento de Astrof{\'\i}sica, Universidad de La Laguna, E-38205, La Laguna, Tenerife, Spain
}

\date{Recieved / Accepted }

\abstract{Sunspots survive on the solar surface for time-scales ranging from days to months. This requires them to be in an equilibrium involving magnetic fields and hydrodynamic forces. Unfortunately, theoretical models of sunspot equilibrium are very simplified as they assume that spots are static and possess a self-similar and axially symmetric magnetic field. These assumptions neglect the role of small scale variations of the magnetic field along the azimuthal direction produced by umbral dots, light bridges, penumbral filaments, and so forth.}{We aim at studying whether sunspot equilibrium is maintained once azimuthal fluctuations in the magnetic field, produced by the sunspot fine structure, are taken into account.}{We apply the FIRTEZ Stokes inversion code to spectropolarimetric observations to infer the magnetic and thermodynamic parameters in two sunspots located at disk center and observed with two different instruments: one observed from the ground with the 1.5-meter German GREGOR Telescope and another with the Japanese spacecraft Hinode. We compare our results with three dimensional radiative magnetohydrodynamic simulations of a sunspot carried out with the MuRAM code.}{We infer clear variations in the gas pressure and density of the plasma directly related to fluctuations in the Lorentz force and associated with the filamentary structure in the penumbra. Similar results are obtained in the umbra despite its lack of observed filamentary structure. Results from the two observed sunspots are in excellent qualitative and quantitative agreement with the numerical simulations.}{Our results indicate that the magnetic topology of sunspots along the azimuthal direction is very close to magnetohydrostatic equilibrium, thereby helping to explain why sunspots are such long-lived structures capable of surviving on the solar surface for days or even full solar rotations.} 

\titlerunning{Lorentz force and sunspot equilibrium}

\authorrunning{Borrero et al.}
\keywords{Sun: sunspots -- Sun: magnetic fields -- Sun: photosphere -- Magnetohydrodynamics
  (MHD) -- Polarization}
\maketitle

\def\kms{~km s$^{-1}$}
\def\deg{^{\circ}}
\def\df{{\rm d}}
\newcommand{\ve}[1]{{\rm\bf {#1}}}
\newcommand{\diff}{{\rm d}}
\newcommand{\Conv}{\mathop{\scalebox{1.5}{\raisebox{-0.2ex}{$\ast$}}}}%
\def\ex{{\bf e_x}}
\def\ez{{\bf e_z}}
\def\ey{{\bf e_y}}
\def\expr{{\bf e_x^\ensuremath{\prime}}}
\def\ezpr{{\bf e_z^\ensuremath{\prime}}}
\def\eypr{{\bf e_y^\ensuremath{\prime}}}
\def\xp{x^\ensuremath{\prime}}
\def\yp{y^\ensuremath{\prime}}
\def\zp{z^\ensuremath{\prime}}
\def\rp{r^\ensuremath{\prime}}
\def\xas{x^{\ast}\!}
\def\yas{y^{\ast}\!}
\def\zas{z^{\ast}\!}
\def\C{\mathcal{C}}
\def\xpp{x^{\prime\prime}}
\def\ypp{y^{\prime\prime}}
\def\zpp{z^{\prime\prime}}

\section{Introduction}
\label{sec:introduction}

Sunspots are the most prominent manifestation of the solar magnetic activity cycle \citep{solanki2003review}. Their inner region, the umbra, is cool and appears dark. The umbra can feature so-called umbral dots and light-bridges, which are intrusions of convective flows that disrupt the otherwise homogeneous magnetic field. The umbra is surrounded by a hotter and brighter halo, the penumbra, that features a very distinct filamentary structure. It is characterized by the presence of Evershed flow channels that are mostly concentrated along horizontal and radially aligned weak magnetic field lines, and that alternate along the azimuthal direction with a more vertical and stronger magnetic field \citep{lites1993pen,solanki1993pen,stanchfield1997pen}.\\

In addition, sunspots harbor a myriad of highly dynamic phenomena that evolve on time scales comparable to the Alv\'en crossing time scales and convective time scales (i.e. minutes to hours in the upper atmosphere). Examples of such phenomena include umbral dots and penumbral grains \citep{sobotka2009}, overturning convection \citep{ichimoto2007pen,scharmer2011pen}, waves \citep{shaun2007atomic,shaun2007pen,johannes2015pen}, Evershed clouds \citep{cabrera2007fts,cabrera2008pen}, and so forth. In spite of this, and due to their long lifetimes, that spans from days to several months, sunspots are thought to be in an equilibrium involving magneto-hydrodynamical forces \citep{rempel2011review}. However, theoretical descriptions of this equilibrium are very simplified. They assume, for instance, that the magnetic field is self-similar and axially symmetric \citep{low1975,low1980a,low1980b,pizzo1986,pizzo1990,jan1994,lena2008}. This assumption completely ignores the large variations of the magnetic field along the azimuthal direction caused by small-scale features such as umbral dots \citep{ortiz2010}, light bridges \citep{falco2016}, penumbral filaments, and so forth. Here we address the long-neglected equilibrium along the azimuthal direction and conclusively demonstrate, using observations and numerical simulations, that in the photosphere both umbra and penumbra are very close to azimuthal magnetohydrostatic equilibrium despite the presence of strong inhomogeneities in the magnetic field. These results provide decisive observational and theoretical support for the idea that sunspots slowly evolve around an equilibrium state and are, in leading order, in magnetohydrostatic equilibrium, thereby helping to explain their long life-spans.\\

\section{Observations}
\label{sec:obs}

The observations analyzed in this work correspond to spectropolarimetric observations in several magnetically sensitive Fe \textsc{i} spectral lines of two sunspots located very close to disk center. The first sunspot was observed from the ground using the GRIS instrument \citep{collados2012gregor} attached to the German 1.5-meter GREGOR solar telescope \citep{schmidt2012gregor}, whereas the second sunspot was observed with the SP instrument \citep{lites2001hinode,ichimoto2007hinode} on-board the Japanese spacecraft Hinode \citep{suematsu2008hinode,tsuneta2008hinode,shimizu2008hinode,kosugi2007hinode}.\\

\begin{table*}
\begin{center}
\caption{Atomic parameters of the spectral lines analyzed in this work}

\begin{tabular}{ccccccccc}
Ion & $\lambda_{0}$\footnotemark[1] & $\chi_{\rm low}$\footnotemark[1] & $\log(gf)$ & Elec.conf\footnotemark[1] & $\sigma$ & $\alpha$ & $g_{\rm eff}$ & Instr. \\
 & [{\AA}] & [eV] & & & & & \\
\hline
Fe \textsc{i} & 6301.5012 & 3.654 & -0.718\footnotemark[4] & $^5$P$_2-{^5}$D$_0$ & 840 & 0.243 & 1.67 & Hinode/SP\\
Fe \textsc{i} & 6302.4936 & 3.686 & -1.165\footnotemark[5] & $^5$P$_1-{^5}$D$_0$ & 856 & 0.241 & 2.5 & Hinode/SP\\
\hline
Fe \textsc{i} & 15648.515 & 5.426 & $-$0.669\footnotemark[2] & ${^7}D_{1}-{^7}D_{1}$ & 975\footnotemark[2] & 0.229\footnotemark[2] & 3.0 & GREGOR/GRIS\\
Fe \textsc{i} & 15662.018 & 5.830 & 0.190\footnotemark[3] & ${^5}F_{5}-{^5}F_{4}$ & 1197\footnotemark[3] & 0.240\footnotemark[3] & 1.5 & GREGOR/GRIS\\
\hline
\end{tabular}
\tablefoot{$\lambda_{0}$ represents the central laboratory wavelength of the spectral line. $\sigma$ and $\alpha$ represent the cross-section (in units of Bohr's radius squared $a_0^2$) and velocity parameter of the atom undergoing the transition, respectively, for collisions with neutral atoms under the ABO theory \citep{abo1,abo2,abo3}. \footnotemark[1]{Values taken from \citet{nave1994}. \footnotemark[2]{Values taken from \citet{borrero2003atomic}}. \footnotemark[3]{Values taken from \citet{shaun2007atomic}}. \footnotemark[4]{Values taken from \citet{bard1991}}. \footnotemark[5]{Values provided by Brian C. Fawcett from the Rutherford Appleton Laboratory (private communication)}\label{table:atomicdata}}}
\end{center}
\end{table*}

For the first sunspot, GREGOR's Infrared Spectrograph (GRIS) was used to record the Stokes vector $\ve{I}^{\rm obs}(x,y,\lambda)=(I,Q,U,V)$ across a 4 nm wide wavelength region centered around 1565 nm and with a wavelength sampling of $\delta_\lambda \approx 40$ m{\AA}~pixel$^{-1}$. This wavelength region was therefore sampled with 1000 spectral points, out of which we selected a 7.6 {\AA} wide region with $N_\lambda=190$ spectral points that included two Fe \textsc{i} spectral lines (see Table~\ref{table:atomicdata}). The spectral line Fe \textsc{i} 15652.8 {\AA} was also recorded but it is not included in our analysis because in the umbra it appears blended with two molecular lines \citep{mathew2003pen} and the FIRTEZ inversion code employed in this work (see Section~\ref{sec:inv}) cannot model them. During data acquisition, five accumulations of 30 ms each were recorded for each modulation state. This results in a total exposure time of 0.6 seconds per slit, yielding a noise level in the polarization signals of $\sigma_q \approx \sigma_u \approx \sigma_v \approx 10^{-3}$ in units of the quiet Sun continuum intensity.\\

In the second sunspot, Hinode's spectropolarimeter (SP) recorded the Stokes vector  $\ve{I}^{\rm obs}(x,y,\lambda)=(I,Q,U,V)$ across a $\approx 0.24$~nm wide wavelength region around 630 nm and with a wavelength sampling of $\delta_\lambda \approx 21.5$ m{\AA}~pixel$^{-1}$. This wavelength region was sampled with a total 112 spectral points including two Fe \textsc{i} spectral lines (see Table~\ref{table:atomicdata}). In these data the total exposure time per slit position is of 4.8 seconds, resulting in similar noise levels in the polarization signals to those of GREGOR data:  $\sigma_q \approx \sigma_u \approx \sigma_v \approx 10^{-3}$ in units of the quiet Sun continuum intensity.\\

The first sunspot is NOAA AR 12049 and was observed on May 3rd, 2015. Different studies of the same sunspot have been presented elsewhere \citep{morten2016,borrero2016pen}. By correlating our images with simultaneous HMI/SDO full-disk continuum images, we estimate that the sunspot center was located at coordinates $(x,y) = (73",-83")$ (measured from disk center). These values correspond a to heliocentric angle of $\Theta=6.5^{\circ}$ ($\mu=0.993$). The image scale is $\delta_x=0.135$"~pixel$^{-1}$ and $\delta_y=0.136$"~pixels$^{-1}$ along the $x$ and $y$ axis, respectively. The scanned region contains $N_x=411$ and $N_y=341$ pixels along each spatial dimension, resulting in a total field of view of $55.5"\times46.4"$. The width of the spectrograph's slit was set to 0.27" (i.e. twice the scanning step). Details about the data calibration in terms of flat-fielding, spectral stray-light correction, fringe removal and wavelength calibration can be found in \citet{borrero2016pen}.\\

The second sunspot is NOAA AR 10933 and was observed on January 6th, 2007. This sunspot has already been analyzed in previous works \citep{morten2013pen}. The sunspot center was located at coordinates $(x,y) = (72",-21")$ (measured from disk center). These values correspond to a heliocentric angle of $\Theta=4.4^{\circ}$ ($\mu=0.997$). The image scale is $\delta_x=0.15$"~pixel$^{-1}$ and $\delta_y=0.16$"~pixels$^{-1}$ along the $x$ and $y$ axis, respectively. The scanned region contains $N_x=350$ and $N_y=300$ pixels along each spatial dimension, resulting in a total field of view of $52.5"\times48.0"$.\\

\section{Stokes inversion}
\label{sec:inv}

In this work we have performed two distinct inversions of each of the datasets described in Sect.~\ref{sec:obs}. The first one is a pixel-wise inversion where each spatial pixel on the solar surface $(x,y)$ is treated independently. The second one is a PSF-coupled inversion employing the Point Spread Functions described in Sect.~\ref{sec:psf}. In some aspects both work identically. Let us describe those first.\\

The inversion of the observed Stokes vector $\ve{I}_{\rm obs}(\lambda,x,y)$ is performed by using an initial atmospheric model described by the following physical parameters: temperature $T(x,y,z)$, magnetic field $\ve{B}(x,y,z)$ and line-of-sight velocity $v_z(x,y,z)$. On the XY plane we use the same number of $(x,y)$ grid cells and spacing as the observations. Along the vertical $z$ direction the atmospheric model is discretized in 128 points with a vertical spacing of $\Delta z=12$ km. The solution to the radiative transfer equation for polarized light from this initial model along the vertical direction yields the synthetic Stokes vector $\ve{I}_{\rm syn}(\lambda,x,y)$, which is compared to the observed ones via a $\chi^2$ merit function. The FIRTEZ code also provides the analytical derivatives of the Stokes vector with respect to the aforementioned physical parameters.\\

It is at this point that the pixel-wise and PSF-coupled inversions differ. On the pixel-wise inversion, the derivatives at location $(x,y)$ enter a Levenberg-Marquardt algorithm \citep{press1986num}, which combined with the Singular Value Decomposition method \citep{golub1965svd}, provides the perturbations in temperature, magnetic field, etc., to be applied to the initial physical parameters so that the next solution to the polarized radiative transfer equation will yield a $\ve{I}_{\rm syn}(\lambda,x,y)$ that is closer to the observed $\ve{I}_{\rm obs}(\lambda,x,y)$ (i.e. $\chi^2$ minimization). The perturbations are only applied at discrete $z$ locations called nodes, with the final $z$ stratification being obtained via interpolation across the 128 vertical grid points. The number of nodes employed in this work is provided in Table~\ref{table:nodes}. The matrices to be inverted, at a given spatial location, are squared matrices with the following number of elements: $[N(T) \times N(B_x) \times N(B_y) \times N(B_z) \times N(v_z)]^2$, where $N(T)$ refers to the number of nodes allowed for the temperature, $N(B_x)$ the number of nodes allowed in the $x$-component of the magnetic field, and so forth. In this way the initial model at location $(x,y)$ is iteratively perturbed until $\chi^2$ reaches a minimum, at which point we assume that the resulting atmospheric model at that location is the correct one.\\

In the PSF-coupled inversion FIRTEZ closely follows the method originally developed by \citet{vannoort2012decon}. It proceeds in a similar fashion as in the pixel-wise inversion up until the $\chi^2$ calculation. The merit function is not obtained by comparing $\ve{I}_{\rm obs}(\lambda,x,y)$ with $\ve{I}_{\rm syn}(\lambda,x,y)$ but rather with $\tilde{\ve{I}}_{\rm syn}(\lambda,x,y)$ which results from the spatial convolution of $\ve{I}_{\rm syn}(\lambda,x,y)$ with the instrumental Point Spread Function $\mathcal{F}(x,y)$. Details about the PSFs used in this work are given in Section~\ref{sec:psf}. Now, the derivatives of the Stokes vector with respect to the physical parameters are coupled horizontally, so that, a perturbation of a physical parameter at location $(\xp,\yp)$ will affect the outgoing Stokes vector at $(x,y)$. Therefore the $\chi^2$ minimization cannot be performed individually for each $(x,y)$ pixel. Instead the minimization must be done at once considering all pixels on the observed region simultaneously. Because of this the square matrices to be inverted are now much larger than in the pixel-wise inversion. The number of elements is now $[N(T) \times N(B_x) \times N(B_y) \times N(B_z) \times N(v_z) \times N_x \times N_y]^2$. For the coupled inversion FIRTEZ does not employ the Singular Value Decomposition method. Instead the coupled matrix is considered to be sparse and its inverse is found via the Biconjugate stabilized gradient method \citep[BiCGTAB ; ][]{vorst1992}.\\

Each of the two described inversions, pixel-wise and PSF-coupled, are first run under the assumption of vertical hydrostatic equilibrium for the gas pressure. Once this is done the magnetic field from the hydrostatic inversion is modified so as to remove the 180$^{\circ}$ ambiguity in $B_x$ and $B_y$ \citep{metcalf2006} via the Non-Potential Field Calculation method \citep{manolis2005}. The resulting atmospheric model is then employed, as an initial guess of an inversion where the gas pressure is obtained under magneto-hydrostatic equilibrium. A more detailed description on the approach followed can be found in \citet{borrero2021firtez}.\\

From all inferred physical parameters by the Stokes inversion, in our study we will focus mainly in the following ones: module of the magnetic field $\|\ve{B}\|$, inclination of the magnetic field with respect to the normal direction to the solar surface: $\gamma = \cos^{-1}(B_z/\|\ve{B}\|)$, gas pressure $P_g$, density $\rho$, and finally also on the Lorentz force $\ve{L} = (4\pi)^{-1} (\nabla \times \ve{B}) \times \ve{B}$. Spatial derivatives of the magnetic field are determined employing fourth-order centered finite differences over an angular array with a constant $\Delta \phi$ step such that $R \Delta \phi$ ($R$ being the arc radius) is approximately equal to the grid size. We note that using second or even sixth order derivatives does not impact results.\\

\begin{table}
\begin{center}
\caption{Number of nodes employed in the Stokes inversion\label{table:nodes}}
\begin{tabular}{cccccc}
Inversion type & T & $v_z$ & $B_x$ & $B_y$ & $B_z$\\
\hline
hydrostatic & 4 & 2 & 2 & 2 & 2\\
\hline
magneto-hydrostatic & 2 & 4 & 4 & 4 & 4\\
\hline
\end{tabular}
\end{center}
\end{table}

\section{Point Spread Functions}
\label{sec:psf}

\begin{figure}[ht!]
\begin{center}
\includegraphics[width=9cm]{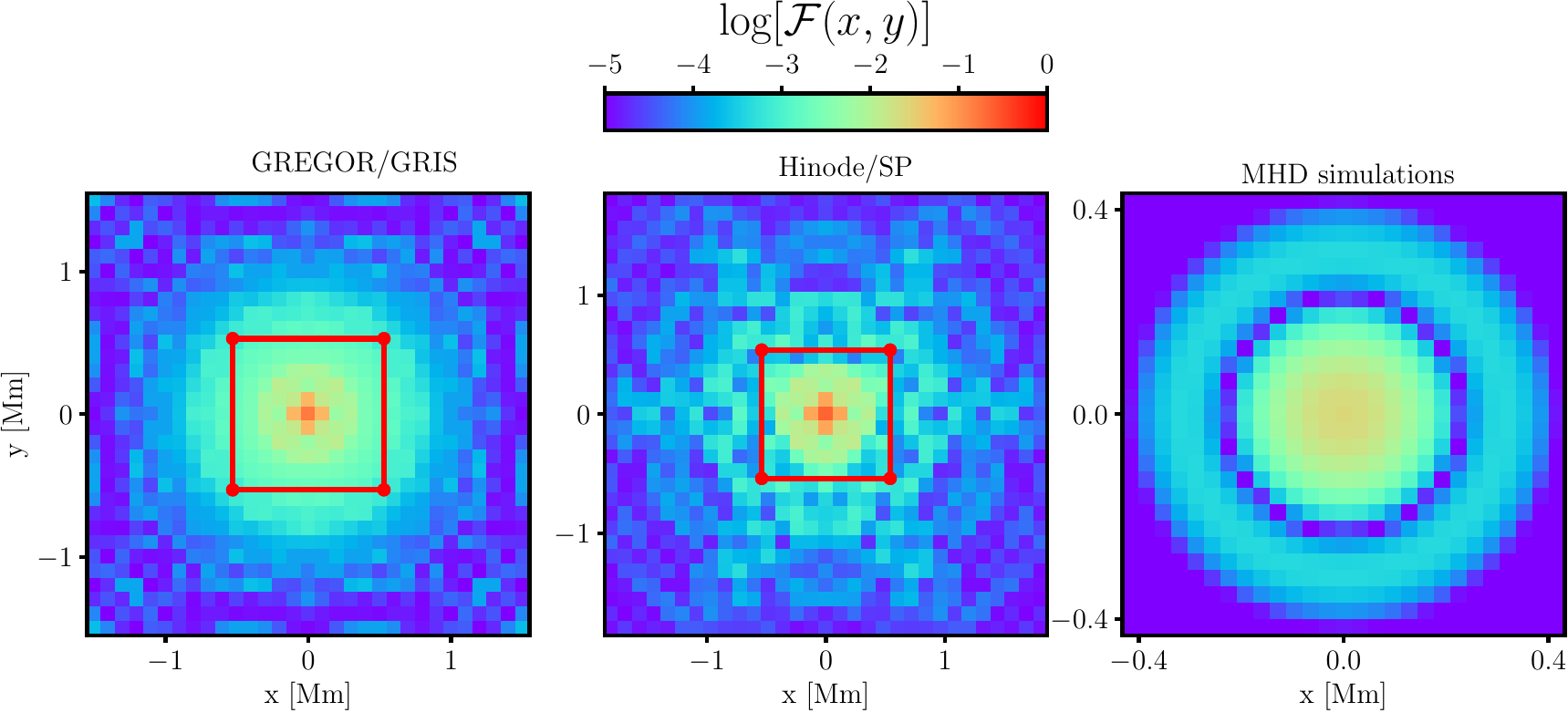}
\end{center}
\caption{Point Spread Functions employed in the data analysis. Left panel: PSF (diffraction limited plus seeing) used in the coupled-inversion of GREGOR/GRIS data. Middle panel: PSF (diffraction limited) used in the coupled-inversion of Hinode/SP data. For both GREGOR and Hinode data the full PSF ($\pm 1.6$~Mm) is used in the synthesis, whereas in the inversion only the square box is considered. Right panel: ideal PSF (Airy function) used to degrade the MHD simulations.\label{fig:psf}}
\end{figure}

PSF-coupled inversions (described in Section~\ref{sec:inv}) were carried out with FIRTEZ employing the Point Spread Functions $\mathcal{F}(x,y)$ shown in Figure~\ref{fig:psf} for GREGOR/GRIS data (left panel) and Hinode/SP data (middle panel). In the latter case, $\mathcal{F}(x,y)$ corresponds to a diffraction-limited PSF for the SOT telescope on-board Hinode at 630 nm, including a 50 cm primary mirror with a 17.2 cm central obscuration caused by the secondary mirror and three 4 cm wide spiders and a 1.5 mm de-focus \citep{danilovic2008psf,elias2024}. In the former case, being GREGOR a ground-based telescope subject to seeing effects, the PSF is modeled as the sum of two different contributions:

\begin{equation}
\mathcal{F}_{\rm gregor}(x,y)= (1-\alpha) \mathcal{F}_{\rm diff}(x,y) + \alpha \mathcal{F}_{\rm seeing}(x,y) \;
\label{eq:psf}
\end{equation}

\noindent where $\mathcal{F}_{\rm diff}(x,y)$ is the diffraction limited PSF for telescope operating at 1565 nm, including a 144 cm primary mirror with a 45 cm central obscuration caused by the secondary mirror and four 2.2 cm wide spiders. The second contribution $\mathcal{F}_{\rm seeing}(x,y)$ is meant to approximately represent the seeing, and it is modeled as a Gaussian function with $\sigma=0.5"$ (or $\approx 375$~km at disk center) and carrying 43 \% of the energy on the detector \citep[$\alpha=0.43$;][]{david2017si}. While originally we had taken a much broader Gaussian with $\sigma=5"$ ($\approx 3750$~km at disk center), the contribution from such Gaussian across the employed area of $\approx 9$~Mm$^{2}$ (see Fig.~\ref{fig:psf}) is essentially constant, leading to a poor convergence of the coupled inversion. A narrower Gaussian allowed the algorithm to converge  much better while still mimicking the effects of the seeing. It is worth noting that, while the actual seeing conditions change during the observations, $\mathcal{F}_{\rm seeing}$ is considered to be time-independent.\\

According to the Rayleigh criterion, if we consider only the diffraction limited contribution to the PSFs, both GREGOR and Hinode data feature a very similar spatial resolution of about 0.3" because, while Hinode's primary mirror is three times smaller, it also operates at approximately three times shorter wavelength.\\

Even though the full PSFs in Figure~\ref{fig:psf} are employed during the synthesis and to compute the merit function $\chi^2$ between the observed and synthetic Stokes profiles ($\ve{I}^{\rm obs}$ and $\ve{I}^{\rm syn}$, respectively), during the inversion (i.e. to determine the derivatives of $\chi^2$ with respect to the physical parameters) only the red squares are considered. In both cases, Hinode and GREGOR, the region enclosed by these red squares contains more than 75 \% of the total amount of light received by a given pixel on the detector. In the case of Hinode the square box covers $\pm 4$~pixels along each spatial dimension around the central pixel, whereas for GREGOR the square box covers $\pm 6$~pixels. For completeness, Fig.~\ref{fig:psf} (right panel) also includes the Point Spread Function employed to degrade the MHD simulations to a spatial resolution comparable to that of GREGOR and Hinode observations (see Section~\ref{sec:simul}).\\

\section{Magneto-hydrodynamic simulations}
\label{sec:simul}

The magneto-hydrodynamic simulations employed in this work were produced with the MuRAM radiative MHD code \citep{voegler2005muram} with the specific adaptations for sunspot simulations as described in \citet{rempel2009mhd,rempel2009mhdb} and \citet{rempel2011mhd,rempel2012mhd}. The code solves the fully compressible MHD equations using a tabulated equation of state that accounts for partial ionization in the solar atmosphere assuming equilibrium ionization \citep[based on the OPAL package by][]{rogers1996opal}. The radiation transport module computes rays in 24 directions using a short characteristics approach as described in \citet{voegler2005muram}. We used 4-band opacities that group spectral lines according to their contribution height in the solar atmosphere.\\

We used a sunspot simulation run started from \citet{rempel2012mhd}, where the magnetic field at the top boundary was forced to be more horizontal with the parameter $\alpha=2$ (see Appendix~\ref{sec:app_sim}). We continued the simulation at a horizontal resolution of 32$\times$32 km and with 16 km vertical resolution while using non-grey radiative transfer with four opacity bins. The very same simulation, but using a different timestamp was employed in \citet{jan2020sunspot}. The data ($P_{\rm g}$, $\rho$, and $\ve{B}$) with the original 32$\times$32 km resolution is convolved with the Point Spread Function presented in Sect.~\ref{sec:psf} (see also rightmost panel in Fig.~\ref{fig:psf}). This PSF consists of an Airy function where the first minimum is located at a distance of 232~km from the peak. This is equivalent to about 0.3" at disk center and therefore very similar to the ideal (i.e. diffraction limited) spatial resolution in GREGOR and Hinode data (see Section~\ref{sec:psf}). We note that, after the application of this PSF, the data from the MHD simulation was further binned from 32$\times$32~ down to 128$\times$128~km so as to have also a comparable spatial sampling as in the observed data. From there, $\gamma$, $\|\ve{B}\|$ and $\ve{L}$ are calculated. The Lorentz force in the MHD simulations is also calculated employing fourth-order centered finite differences to determine the spatial derivatives of the magnetic field (see Section~\ref{sec:inv}). A summary of the physical properties in the simulations is displayed in Figure~\ref{fig:inv} (third column). Further details about the initial and boundary conditions are given in Appendix~\ref{sec:app_sim}.\\

\section{Results}
\label{sec:results}

\begin{figure*}[ht!]
\begin{center}
\begin{tabular}{ccc}
\includegraphics[width=5cm]{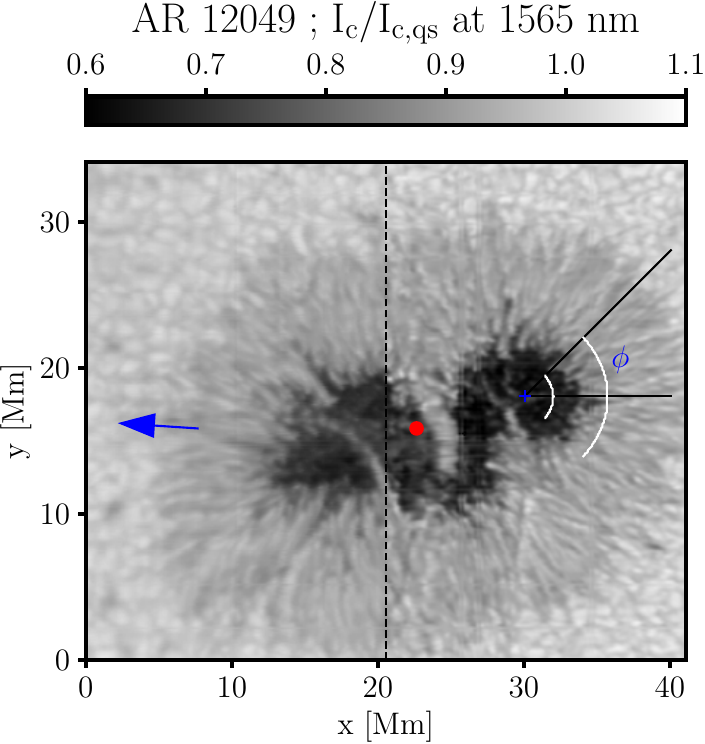} &
\includegraphics[width=5cm]{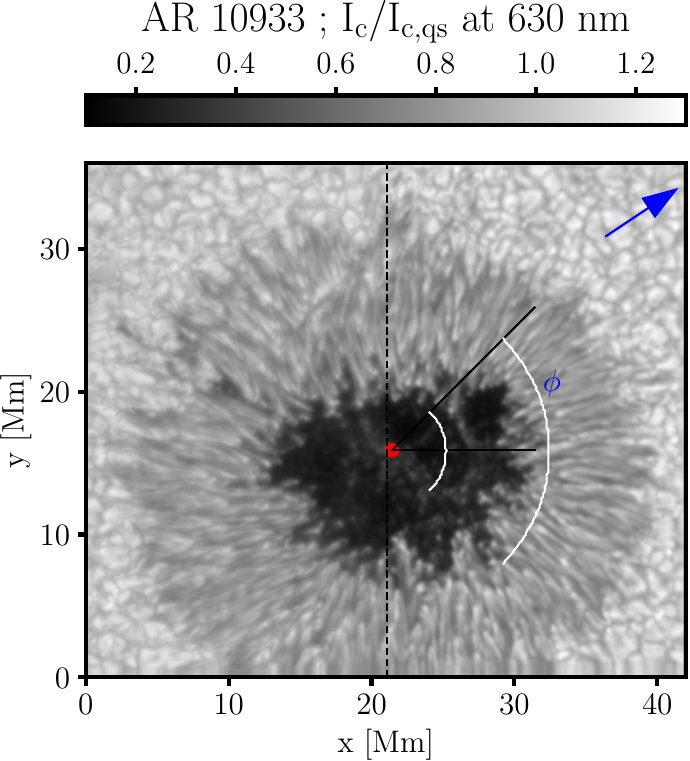} &
\includegraphics[width=5cm]{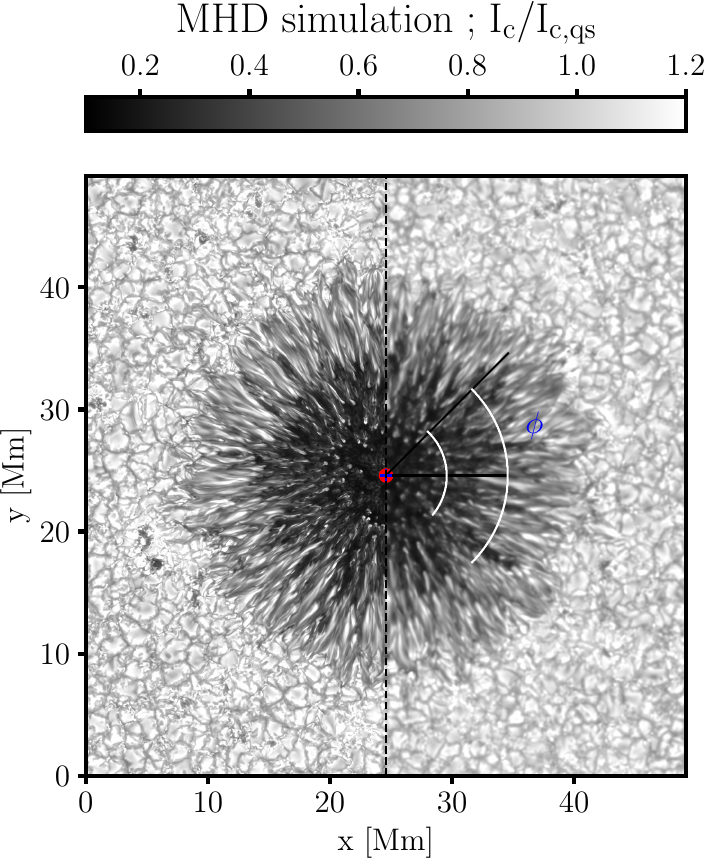} \\
\includegraphics[width=5cm]{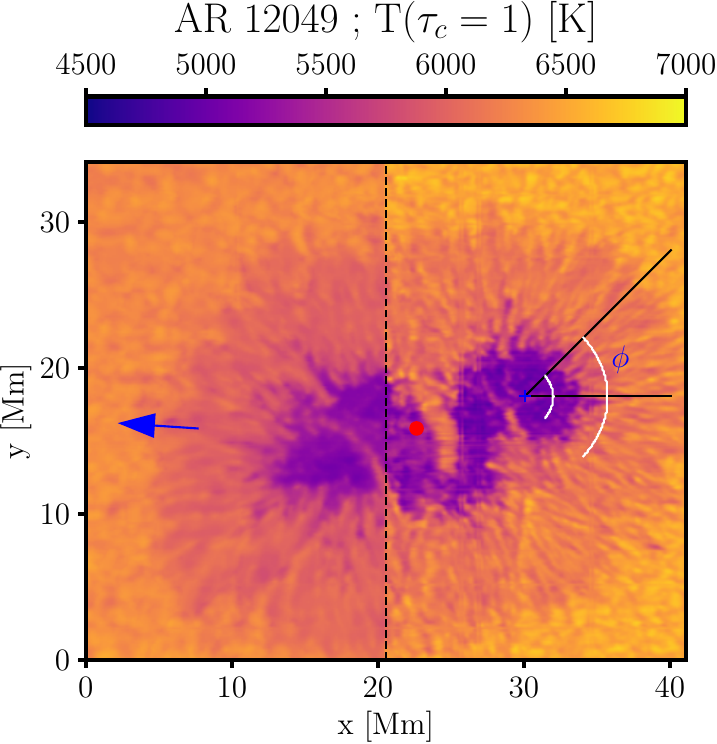} &
\includegraphics[width=5cm]{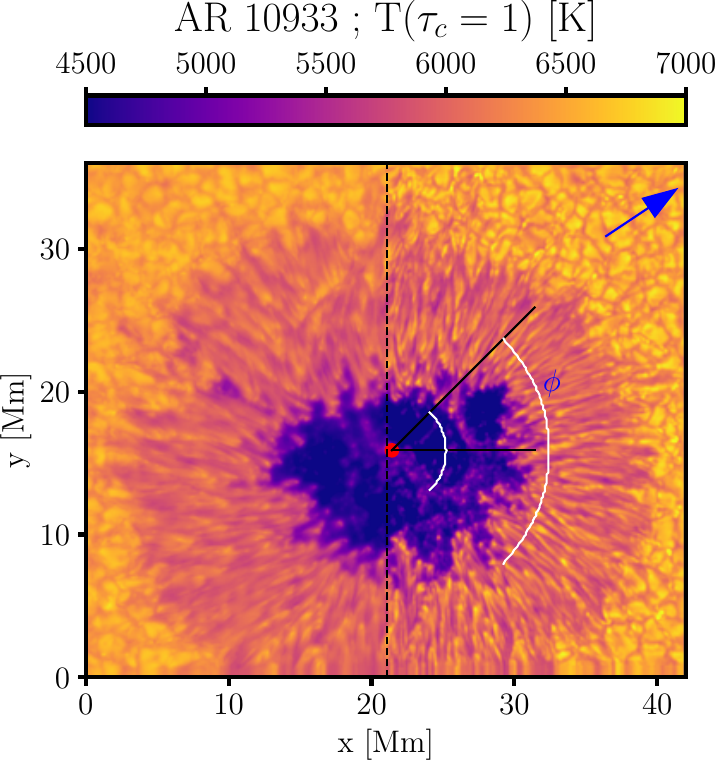} &
\includegraphics[width=5cm]{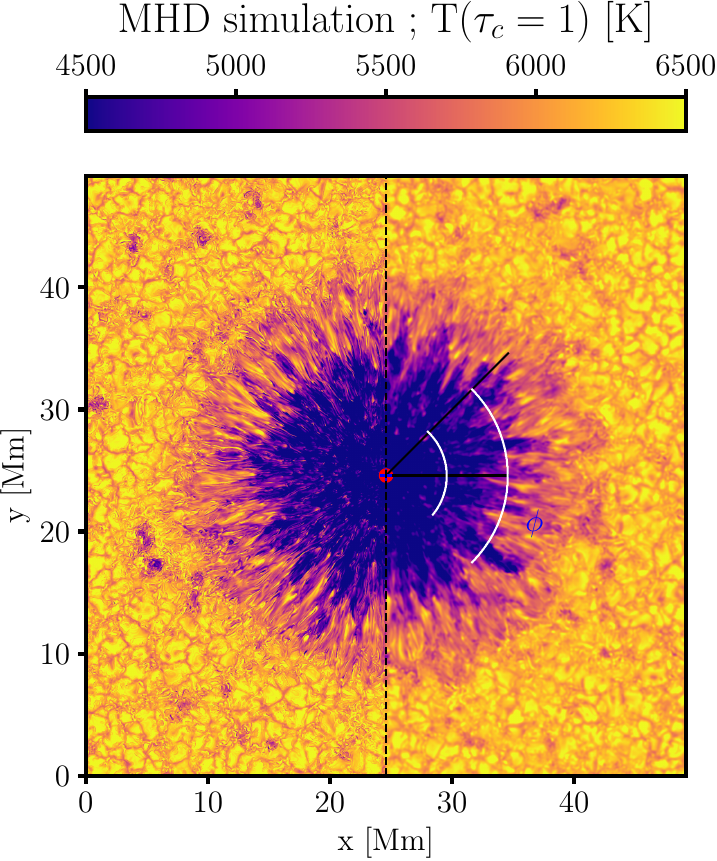} \\
\includegraphics[width=5cm]{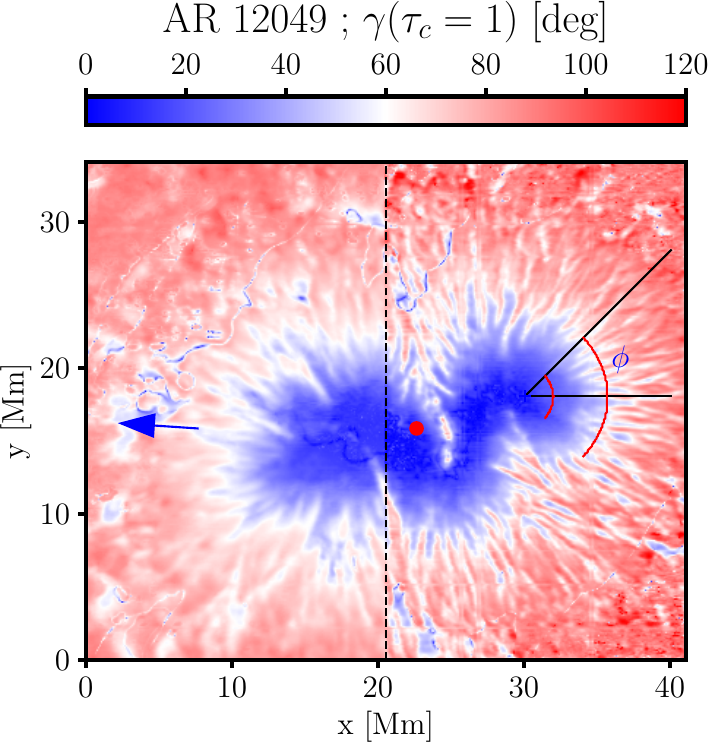} &
\includegraphics[width=5cm]{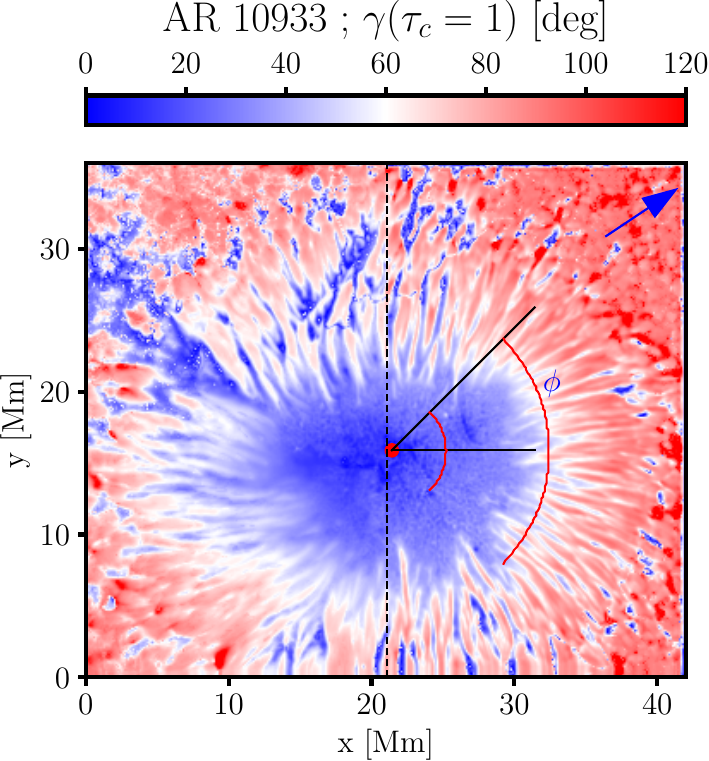} &
\includegraphics[width=5cm]{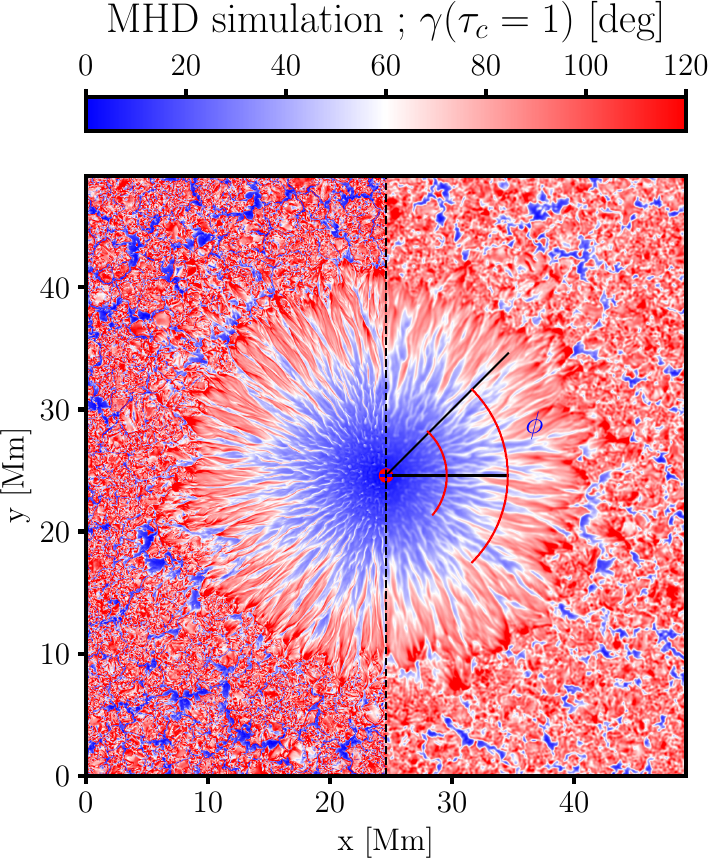} \\
\end{tabular}
\caption{Summary of analyzed data and inversion results. Top row: continuum intensity normalized to the average quiet Sun intensity $I_c/I_{\rm c,qs}$. Middle row: temperature $T$. Bottom row: inclination of the magnetic field $\gamma$ with respect to the normal vector to the solar surface. All values are taken at the continuum optical depth $\tau_{\rm c}=1$. Left column: sunspot observed with the ground telescope GREGOR and data inverted with the FIRTEZ code. Middle column: same as before but for the sunspot recorded with the Hinode satellite. Right column: simulated sunspot with the MuRAM code. Each panel is split on two: the left side of the first two columns display the results of the pixel-wise Stokes inversion, whereas the right side presents the results of the PSF-coupled inversion. On the third column, the left side shows the simulation with the original horizontal resolution of 32 km $\times$ 32 km, while the right side shows the simulation convolved and resampled to 128 km $\times$ 128 in order to match the approximate resolution and sampling of the GREGOR and Hinode data. Each figure contains the direction of the center of the solar disk (blue arrow) and displays two azimuthal paths with varying angle $\phi$ at constant radial distance from the spots' center: the outermost arc is in the penumbra whereas the innermost arc lies in the umbra.\label{fig:inv}}
\end{center}
\end{figure*}

Figure~\ref{fig:inv} illustrates some of the physical parameters inferred from the Stokes inversions and from the MHD simulations. As expected \citep[see][]{vannoort2012decon}, the PSF-coupled inversion (right-side panels on the first two columns) display a larger variation of the physical parameters than the pixel-wise inversion (left-side panels on the first two columns). It is worth mentioning that this the first time that this method is successfully applied to ground-based spectropolarimetric data.\\

Two azimuthal cuts or arcs (i.e along the coordinate $\phi$) are also indicated in Figure~\ref{fig:inv}. The outermost and innermost arcs are located in the penumbra and umbra, respectively. Along the azimuthal direction in the penumbra (outermost arc indicated by the $\phi$ coordinate) there are alternating regions of large inclination ($\gamma \approx 90^{\circ}$; field lines contained within the solar surface) and regions of much lower inclination ($\gamma \approx 40^{\circ}$; field lines more perpendicular to the solar surface). These large variations in the inclination of the magnetic field are produced by the filamentary structure of the penumbra \citep{stanchfield1997pen,martinez1997pen,mathew2003pen,borrero2008penb}. Across the umbra (innermost arc indicated by the $\phi$ coordinate) the inclination $\gamma$ features a much more homogeneous distribution.\\

\begin{figure*}[ht!]
\begin{center}
\begin{tabular}{cc}
\includegraphics[width=8cm]{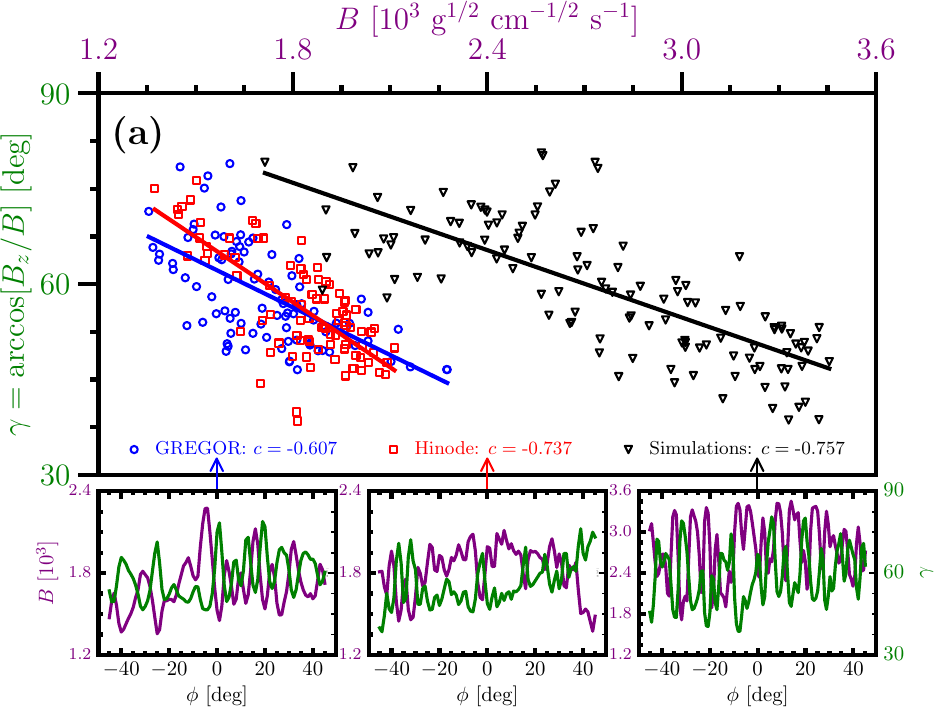} &
\includegraphics[width=8cm]{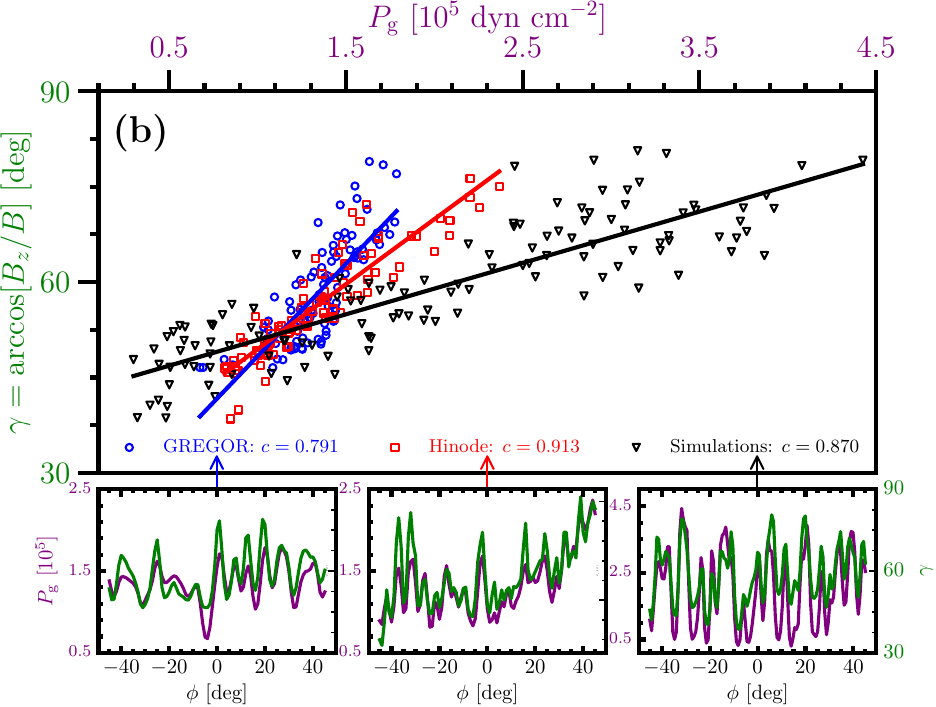} \\
\includegraphics[width=8cm]{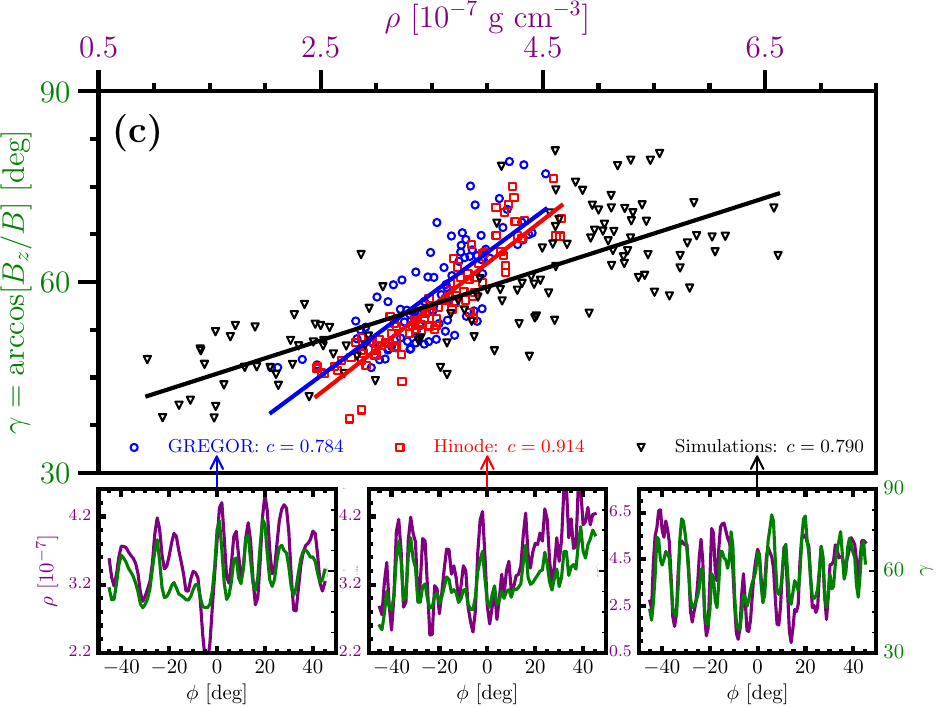} &
\includegraphics[width=8cm]{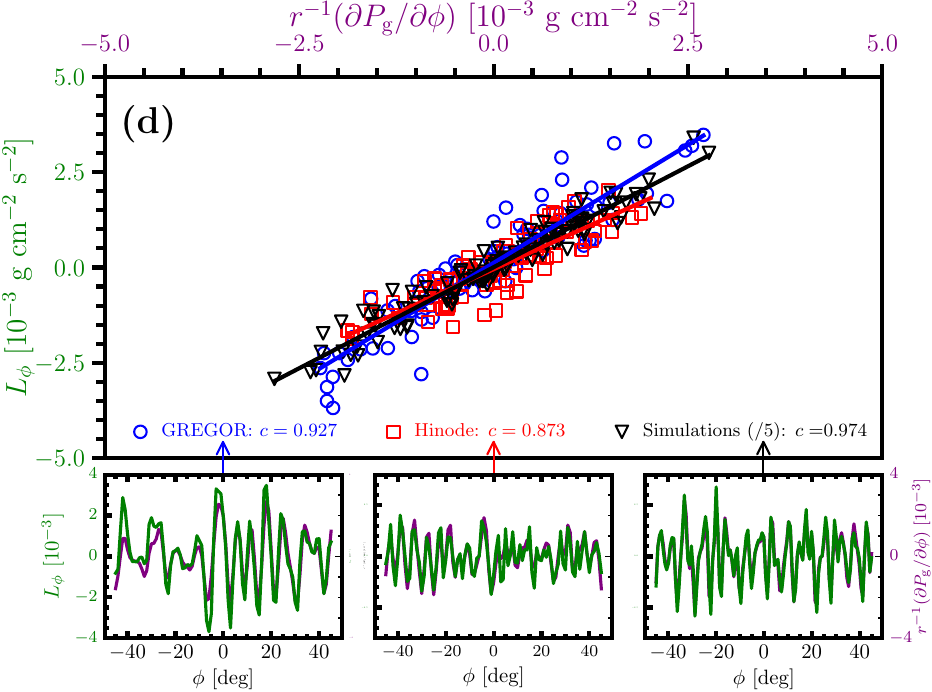} 
\end{tabular}
\caption{Scatter plots of the physical quantities along the outer-most azimuthal arc (i.e. penumbra) for GREGOR data (red squares), Hinode data (blue circles) and MHD simulation (black triangles). Pearson correlations coefficients $c$ for each of these data are provided in the legend. Linear fits to the data are shown in thick color solid lines. Upper-left: $\gamma-\|\ve{B}\|$; Upper-right: $\gamma-P_{\rm g}$; Lower-left: $\gamma-\rho$; Lower-right: $L_\phi-r^{-1}\partial P_{\rm g}/\partial \phi$. Each panel also includes three smaller subpanels where the variations of the physical quantities along $\phi$ are displayed in purple and green colors. These subpanels are ordered as: GREGOR (left), Hinode (middle), and MHD simulations (right). This is also indicated by the arrows under the correlation coefficients for each data source. We note that, the values of the $\phi$ components of the Lorentz force and of the gas pressure gradient in the simulations are divided by five.\label{fig:phi}}
\end{center}
\end{figure*}

Figure~\ref{fig:phi} shows scatter plots of the following pairs of physical quantities along the azimuthal direction ($\phi$-coordinate) in the penumbra (i.e. outermost arc in Fig.~\ref{fig:inv}): a) $\gamma-\|\ve{B}\|$ (upper-left), b) $\gamma-P_{\rm g}$ (upper-right), c) $\gamma-\rho$ (lower-left), and d) $L_\phi- r^{-1}\partial P_{\rm g}/\partial \phi$ (lower-right). In this figure correlations are indicated by $c$ and are provided separately for each of the two observed sunspots as well as for the degraded MHD simulations. We note that, for the original (i.e. undegraded) simulation data, the correlations are even higher than the ones presented here in spite of featuring variations in the physical parameters that are unresolved in the degraded simulation data. Below each scatter plot, three minipanels display the physical quantities along the azimuthal direction. All physical parameters are taken at a constant height $z^{*}$ equal to the average height along $\phi$ where the continuum optical depth is unity: $\tau_c=1$. A similar plot to Fig.~\ref{fig:phi} but for the umbra (i.e innermost arc in Fig.~\ref{fig:inv}) is provided in Section~\ref{sec:umbra} and Fig.~\ref{fig:phi_umbra}.\\

Panel a) in Fig.~\ref{fig:phi} shows a clear anti-correlation between $\gamma$ and $\|\ve{B}\|$. It highlights the well-known uncombed structure of the penumbral magnetic field \citep{solanki1993pen}, where regions of strong and vertical magnetic fields ($\gamma$ low; $\|\ve{B}\|$ large) called spines, alternate along the $\phi$ coordinate with regions of weaker and more horizontal magnetic fields ($\gamma$ large; $\|\ve{B}\|$ low) called intraspines \citep{title1993,lites1993,martinez1997pen}. Panels b) and c) show a clear correlation in regions where the inclination of the magnetic field $\gamma$ is larger (i.e. intraspines) and regions of enhanced gas pressure $P_{\rm g}$ and density $\rho$. These curves indicate that the penumbral intraspines possess an excess gas pressure and density compared to the spines. This decade-long theoretical prediction \citep{spruit2006gap,scharmer2006gap,borrero2007pen} is observationally confirmed here for the first time.\\

Panel d) in Fig.~\ref{fig:phi} shows that the gas pressure and density fluctuations, associated with the intraspine/spine structure of the penumbral magnetic field and highlighted in panel a), are directly related to fluctuations in the azimuthal component of the Lorentz force. This is demonstrated not only by the high degree of correlation between $L_\phi$ and $r^{-1}\partial P_{\rm g}/\partial\phi$, but also by the fact that the slope of the scatter plot is unity, thus indicating that:

\begin{equation}
L_\phi = r^{-1}\partial P_{\rm g}/\partial\phi \;\;\,
\label{eq:mhs_phi}
\end{equation}

which corresponds to the $\phi$ component (in cylindrical coordinates) of the magnetohydrostatic equilibrium equation:\\

\begin{equation}
\nabla P_{\rm g} = \rho \ve{g} + \ve{L} \;\;,
\label{eq:mhs_all}
\end{equation}

where $\ve{L}$ has already been defined as the Lorentz force.\\

Our combined results prove that, along the azimuthal $\phi$ direction, penumbral intraspines/spines are in almost perfect magnetohydrostatic equilibrium and that neither the time derivative of the velocity $\partial \ve{v} / \partial t$, the velocity advective term $(\ve{v}\cdot\nabla) \ve{v}$, nor the viscosity term $\nabla \hat{\tau}$, where $\hat{\tau}$ is the viscous stress tensor, play a significant role in the force balance. Interestingly, in the numerical simulations, the velocity advective term associated to the Evershed flow does indeed play an important role in the force balance along the radial direction \citep{rempel2012mhd}, meaning that magnetohydrostationary equilibrium is more adequate in the radial coordinate.\\

It should be noted that the almost perfect magnetohydrostatic equilibrium along the azimuthal direction (Eq.~\ref{eq:mhs_phi}) is found in the observations at short and intermediate radial distances from the center of the sunspots: $r/R_{\rm spot} \in [0.1,0.7]$ (where $R_{\rm spot}$ is the radius of the sunspot) and in a height region restricted around $\pm 150$ km around the continuum $\tau_{\rm c}=1$ (see Section~\ref{sec:mhs} for more details). In the MHD simulations the magnetohydrostatic equilibrium is verified over a similar range of radial distances but over a much broader vertical region of about 400 km above $\tau_{\rm c}=1$.\\ 

\subsection{Contribution of the magnetic tension and magnetic pressure gradient in the Lorentz force}
\label{sec:lor}

\begin{figure}[ht!]
\begin{center}
\includegraphics[width=9cm]{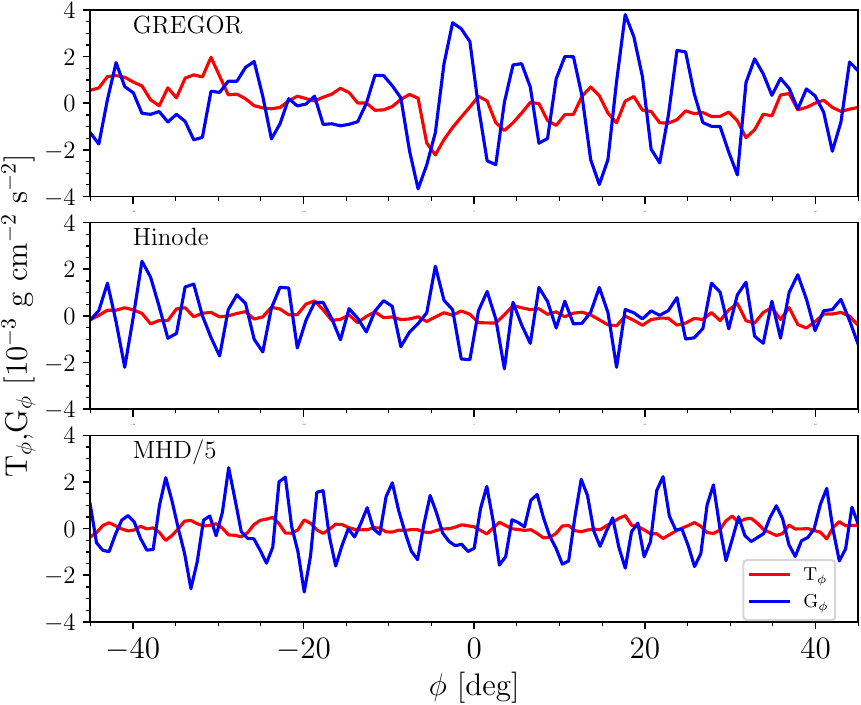}
\caption{Decomposition of the Lorentz force into magnetic pressure gradient and magnetic tension. Variations of the magnetic pressure gradient $G_\phi$ (blue) and magnetic tension $T_\phi$ (red) along the azimuthal coordinate $\phi$ indicated in Fig.~\ref{fig:inv} (outermost arc) and inferred from GREGOR (top) and Hinode (middle) observations, as well as from 3D MHD simulations (bottom). Values for the latter are divided by a factor of five.\label{fig:deconstruct_fl}}
\end{center}
\end{figure}

We have already established that the azimuthal component of the Lorentz force $L_\phi$ has a one-to-one relationship to the pressure and density fluctuations along this direction seen both in observations and simulations (i.e. Eq.~\ref{eq:mhs_phi} is verified). In order to study this in more detail, we have decomposed $L_\phi$ into the contribution from the magnetic tension $T_\phi$ and from magnetic pressure gradient $G_\phi$:

\begin{eqnarray}
L_\phi & = & \underbrace{(4\pi)^{-1} ({\bf B} \cdot \nabla) B_\phi}_{T_\phi}-\underbrace{\frac{1}{8\pi r} \frac{\partial \|{\bf B}\|^2}{\partial \phi}}_{G_\phi}
\end{eqnarray}

\noindent where $\ve{e}_\phi$ is the unit vector in the azimuthal direction. The resulting azimuthal variations of the magnetic tension and magnetic pressure gradient are shown in Figure~\ref{fig:deconstruct_fl}. As it can be seen the similarities between GREGOR and Hinode observations, with the MHD simulations are remarkable. In all cases, the magnetic pressure gradient $G_\phi$ dominates over the magnetic tension $T_\phi$, with the former reaching its maximum right at the boundary between spines and intraspines, and the latter peaking at the center of spines/intraspines respectively.\\

\subsection{Gas pressure fluctuations: observations vs MHD simulations}
\label{sec:bfield}

In Figure~\ref{fig:phi} (panel d) we showed the results of the three dimensional MHD numerical simulations but dividing $L_\phi$ and $r^{-1} \partial P_{\rm g} / \partial \phi$ by a factor of five so that the points of the scatter plot appear on the same scale as the ones from the GREGOR and Hinode observations. The same factor is applied to the decomposition of the Lorentz force into the magnetic tension and gas pressure gradient contributions in Figure~\ref{fig:deconstruct_fl} (bottom panel). The reason for the larger values of the Lorentz force and of the gas pressure variations along the azimuthal direction in the MHD simulations is to be found in Figure~\ref{fig:phi} (panel a) where it is shown that the MHD simulations have a much larger range of variations of the magnetic field $\|\ve{B}\| \in [2000,3500]$~Gauss compared to the observations in GREGOR and Hinode data, where $\|\ve{B}\| \in [1300,1900]$. In order to demonstrate this, we take the force balance equation along the azimuthal direction (Eq.~\ref{eq:mhs_phi}) and consider, as seen in Sect.~\ref{sec:lor}, that the main contributor to the $\phi$ component (azimuthal direction) of the Lorentz force is the magnetic pressure gradient along the same direction. With this, we can write:

\begin{equation}
\frac{1}{r} \frac{\partial P_{\rm g}}{\partial \phi} \simeq G_\phi = -\frac{1}{8\pi} (\nabla \|\ve{B}\|^2) \ve{e}_\phi = -\frac{1}{8\pi r} \frac{\partial B^2}{\partial \phi}
\end{equation}

By transforming the derivatives into finite differences we can write:\\

\begin{equation}
\frac{\Delta P_{\rm g, mhd}}{\Delta P_{\rm g, obs}} \simeq \frac{\Delta B^2_{\rm mhd}}{\Delta B^2_{\rm obs}} \approx \frac{3500^2-2000^2}{1900^2-1300^2} \approx 4.3
\end{equation}

This demonstrates that the reason the $\phi$ component of the Lorentz force and hence also the variations of the gas pressure long the azimuthal direction in the MHD simulations is larger than in the observations is because of the larger difference in the magnetic field of the spines and intraspines in the MHD simulations.\\

In the original (i.e. undegraded) simulation data (see left-side panels on the left-most column in Fig.~\ref{fig:inv}), the mean azimuthal distance between consecutive peaks in $P_{\rm g}$ decreases with respect to the degraded simulation data, thereby further increasing the values of $L_\phi$ and $r^{-1} \partial P_{\rm g} / \partial \phi$.\\

\subsection{Results for the umbra}
\label{sec:umbra}

\begin{figure*}[ht!]
\begin{center}
\begin{tabular}{cc}
\includegraphics[width=8cm]{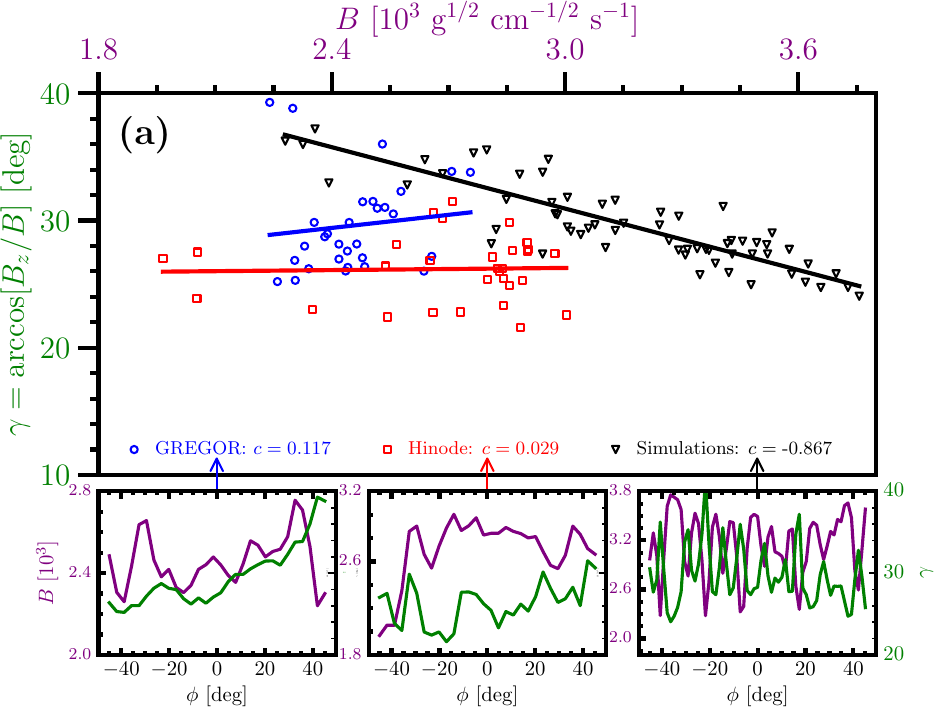} &
\includegraphics[width=8cm]{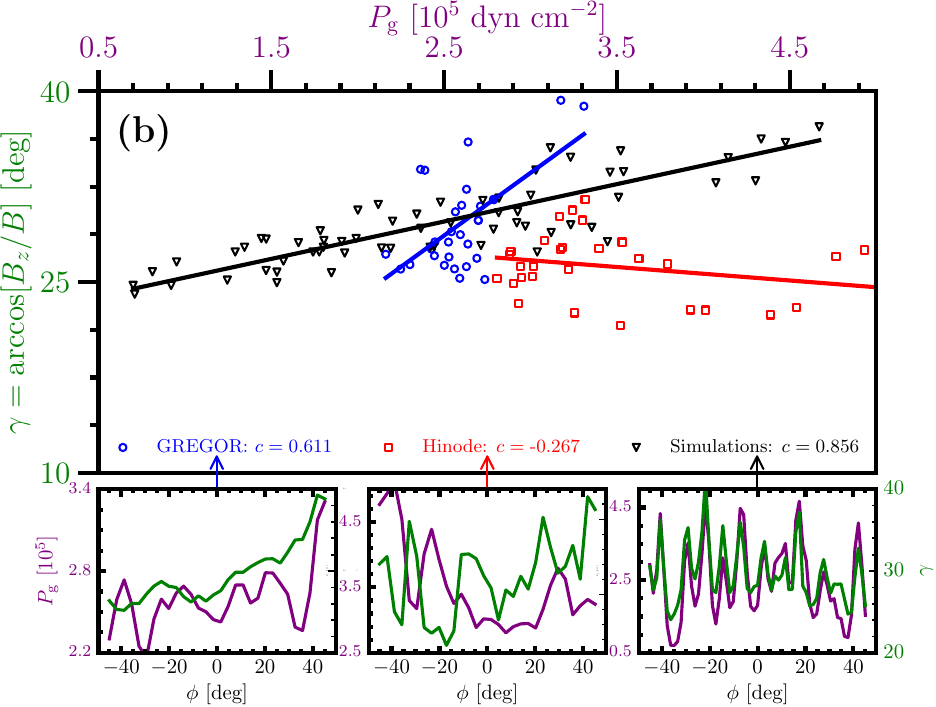} \\
\includegraphics[width=8cm]{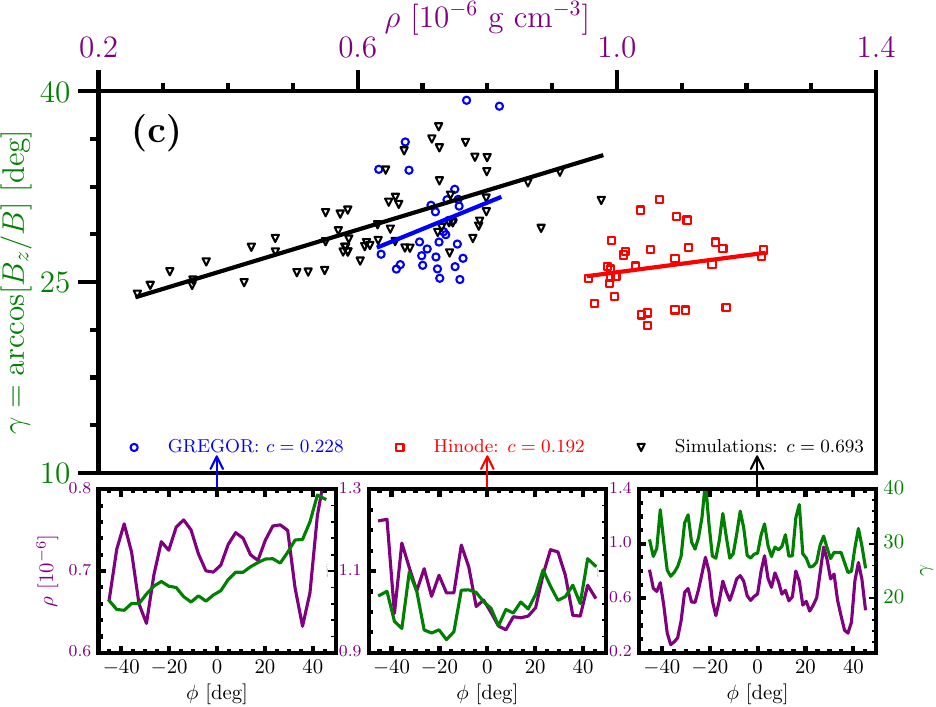} &
\includegraphics[width=8cm]{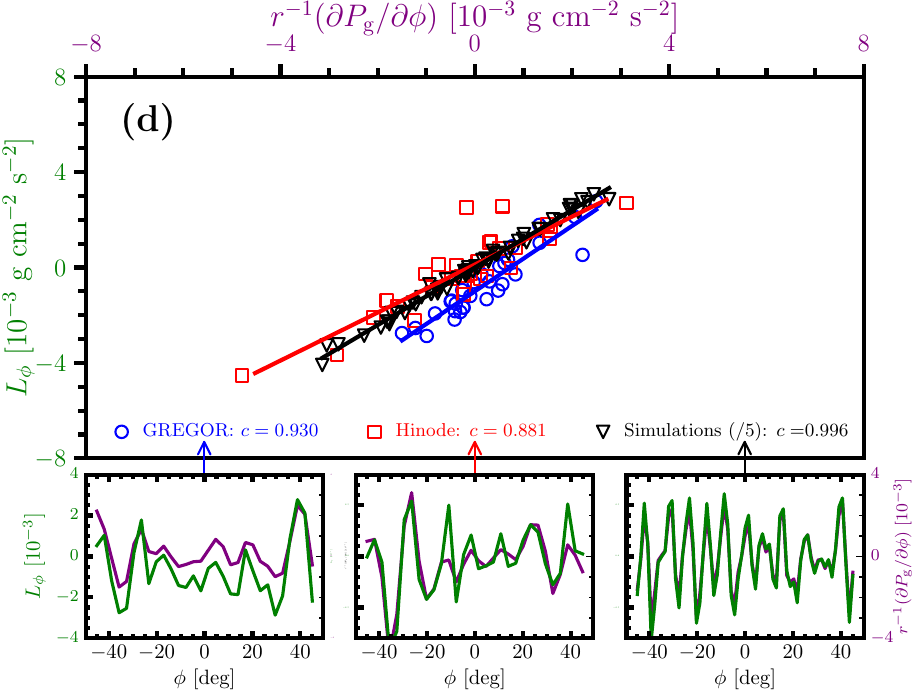} 
\end{tabular}
\caption{Same as Figure~\ref{fig:phi} but in the umbra (i.e. innermost arc in Fig.~\ref{fig:inv}).
\label{fig:phi_umbra}}
\end{center}
\end{figure*}

So far we have mostly focused on the azimuthal equilibrium of the penumbra. Following the same procedure previously described, we can also study whether the umbra satisfies the MHS equilibrium given by: $L_\phi = r^{-1} \partial P_{\rm g} / \partial \phi$. To this end, we present in Figure~\ref{fig:phi_umbra} the same study as in Figure~\ref{fig:phi} but along the innermost arc (i.e. umbral arc; Fig.~\ref{fig:inv}). In panel a) of Figure~\ref{fig:phi_umbra} we notice that, in the observations (red squares for Hinode and blue circles for GREGOR), the strong anti-correlation between $\gamma$-$\|\ve{B}\|$ is not present anymore. This was to be expected because the spine/intraspine structure of the penumbral magnetic field does not appear in the umbra. This is further seen in panels b) and c) where positive correlations between $\gamma$-$P_{\rm g}$ and $\gamma$-$\rho$, respectively, are not inferred in neither Hinode nor GREGOR observations.\\

It is important to notice that while in the penumbra the physical properties inferred from GREGOR and Hinode data were quite similar (see Fig.~\ref{fig:phi}), in the umbra there are large differences between the two observed sunspots. This is particularly the case of the field strength obtained from the Stokes inversion (see a-panel in Fig.~\ref{fig:phi_umbra}) with Hinode showing a larger variation in the field strength. This is likely caused by the presence of umbral dots along the innermost arc in Hinode data. The larger variation in the field strength along the $\phi$ coordinate then translates into a larger variation of the gas pressure and density once MHS equilibrium is taken into account.\\

In the MHD simulations, correlations and anti-correlations between the aforementioned physical parameters are still present, hinting at a clear presence of spine/intrapine magnetic fields in the simulated umbra. This was also to be expected if we attend to the innermost arc in Fig.~\ref{fig:inv} (third column). It is clear from continuum intensity and inclination maps of the MHD simulations that penumbral filaments still exist within the umbra almost up to the very center of the sunspot. This effect appears because of the time stamp, during the evolution of the sunspot, chosen for our analysis. Initially, the convection in the outer umbra shows a typical coffee-bean pattern such as the one reported in \citet{manfred2006umbra}. This pattern closely resembles the observations. However, as time passes and the cooling front goes deeper, the convection within the outer umbra shows no longer coffee beans but is more penumbra-like, with radially aligned filaments, albeit less strong and transporting less energy.\\

Regardless of the differences between the observed and simulated umbra, the most important fact to notice is panel d) in Fig.~\ref{fig:phi_umbra}, where it is still the case that in both observed and simulated umbra, the magnetohydrostatic equilibrium given by Eq.~\ref{eq:mhs_phi} is strongly satisfied. This demonstrates that along the azimuthal direction, the umbra is also in magnetohydrostatic equilibrium.\\

\subsection{Results with vertical hydrostatic equilibrium}\label{sec:hydro}

As described in Section~\ref{sec:inv}, as part of the inversion procedure and before the inversion with magnetohydrostatic (MHS) constraints is applied, a first inversion under vertical hydrostatic equilibrium (HE) is carried out. To study whether the inversion under vertical hydrostatic equilibrium yields similar results to those under magnetohydrostatic equilibrium, we present in Figure~\ref{fig:phi_hydro} the same results as in Fig.~\ref{fig:phi} but using HE instead of MHS. As it can be seen, panel a) in Figure~\ref{fig:phi_hydro} still features the widely known spine/intraspine structure of the penumbral magnetic field, whereby more inclined magnetic field lines ($\gamma \rightarrow 90^{\circ}$) are weaker than the more vertical ($\gamma \rightarrow 30^{\circ}$) magnetic field lines. However, the correlations between the magnetic field inclination $\gamma$ with the gas pressure and density (panels b and c, respectively) as well as the correlations between the gas pressure variation and the Lorentz force along $\phi$, are completely lost under the assumption of vertical hydrostatic equilibrium. This result highlights that the conclusions presented in this paper would not have been possible unless MHS is considered.\\

\subsection{Results without coupled inversion}\label{sec:res_uncoupled}

In Figure~\ref{fig:phi_pixelwise} we show the same results as those presented in Figure~\ref{fig:phi} but performing the MHS inversion using the pixel-wise inversion instead of the PSF-coupled inversion (see Sect.~\ref{sec:inv}). By comparing these two figures we conclude that, although the PSF-coupled inversion is not strictly compulsory to obtain a correlation between the azimuthal component of the Lorentz force $L_\phi$ and the derivative of the gas pressure along the azimuthal direction (panel d) it certainly helps to improve the correlations of the field strength $\|\ve{B}\|$ (panel a), gas pressure $P_{\rm g}$ (panel b), and density $\rho$ (panel c) with the inclination of the magnetic field $\gamma$ with respect to the normal vector to the solar surface. The effects of including the PSF in the inversion are also more noticeable in the GREGOR data (ground-based) than in the Hinode data (satellite). This occurs in spite of our limited knowledge of GREGOR's PSF, in particular the lack of a detailed model for the seeing (see Sect.~\ref{sec:psf}).\\ 

\section{Conclusions }
\label{sec:conclusions}

Although simplified theoretical magnetohydrostatic models of sunspots have been proposed in the past, those models have so far ignored the strong variations of the magnetic field along the azimuthal direction. These variations are caused not only by the filamentary structure of the penumbra but also by inhomogeneities in the umbral magnetic field caused by umbral dots and light bridges. Here we have demonstrated, using spectropolarimetric observations from both the ground and space as well as magnetohydrodynamic simulations, that magnetohydrostatic equilibrium is maintained even if these azimuthal variations in the magnetic field are taken into account.\\

It is important to bear in mind that not all dynamic phenomena occurring at small scales in sunspots (see i.e. Section~\ref{sec:introduction}) are resolved at the spatial resolution of our observations, with many of them being highly relevant for the energy transport within the sunspot \citep{rempel2021review}. Consequently, the magnetohydrostatic equilibrium in the azimuthal direction unearthed by the present study might not apply at spatial scales smaller than those of our observations \citep[i.e. $\le 100$~km][]{schlichenmaier2016}. What can be stated however, is that at those spatial scales, these aforementioned dynamic phenomena must reach a stationary state whereby the overarching magnetic structure of the sunspot remains close to magnetohydrostatic \citep{rempel2011review}. On the other hand, we note that the undegraded magnetohydrodynamic simulation, featuring a horizontal resolution of $32$~km, verify the magnetohydrostatic equilibrium along the azimuthal direction with even higher accuracy than the same simulations after degradation to the spatial resolution of the observations.\\

Notwithstanding their almost perfect magnetohydrostatic equilibrium, perturbations from this equilibrium, such as fluting instability \citep{meyer1977} or advective transport of magnetic flux via moving magnetic features \citep{valentin2002decay,kubo2008mmf2,kubo2008mmf1}, will eventually break apart or make the sunspot decay till its eventual disappearance over long time scales.\\

The results presented in this work also highlight the capabilities of the FIRTEZ Stokes inversion code to infer physical parameters such as gas pressure, density, electric currents and Lorentz force that, until now, could not be accurately determined. Gaining access to these physical parameters opens a new path to investigate the thermodynamic state and magnetic structure of the plasma in the solar atmosphere.\\

In this work we have not studied the equilibrium of sunspots neither along the radial nor the vertical direction $(r,z)$. There are several important reasons for this. On the one hand these have already been addressed by theoretical models \citep{low1975,low1980a,low1980b,pizzo1986,pizzo1990}. On the other hand, the magnetic field along these two directions varies much more slowly than along $\phi$ \citep{borrero2011review} and therefore the Lorentz force will likely play a less important role there. Finally, as shown by \citet[][see Fig.~22]{rempel2012mhd}, along the radial direction in the penumbra, the velocity advective term $(\ve{v}\cdot\nabla) \ve{v}$ associated to the Evershed flow contributes significantly to the equilibrium. This particular term cannot be so far considered by the FIRTEZ inversion code and therefore it is not possible at this point to address the equilibrium along the radial direction employing spectropolarimetric observations.\\

\begin{acknowledgements}
The development of the FIRTEZ inversion code is funded by two grants from the Deutsche Forschung Gemeinschaft (DFG): projects 321818926 and 538773352. Adur Pastor acknowledges support from the Swedish Research Council (grant 2023-03313). This project has been funded by the European Union through the European Research Council (ERC) under the Horizon Europe program (MAGHEAT, grant agreement 101088184). Markus Schmassmann is supported by the Czech-German common grant, funded by the Czech Science Foundation under the project 23-07633K and by the Deutsche Forschung Gemeinschaft (DFG) under the project 511508209. Manuel Collados acknowledges financial support from Ministerio de Ciencia e Innovaci\'on and the European Regional Development Fund through grant PID2021-127487NB-I00. Matthias Rempel would like to acknowledge high-performance computing support from Cheyenne (doi:10.5065/D6RX99HX) provided by NSF NCAR's Computational and Information Systems Laboratory (CISL), sponsored by the U.S. National Science Foundation. This material is based upon work supported by the NSF National Center for Atmospheric Research, which is a major facility sponsored by the U.S. National Science Foundation under Cooperative Agreement No. 1852977.  The 1.5-m GREGOR solar telescope was built by a German consortium under the leadership of the Institut f\"ur Sonnenphysik in Freiburg with the Leibniz-Institut f\"ur Astrophysik Potsdam, the Institut f\"ur Astrophysik G\"ottingen, and the Max-Planck-Institut f\"ur Sonnensystemforschung in G\"ottingen as partners, and with contributions by the Instituto de Astrof{\'\i}sica de Canarias and the Astronomical Institute of the Academy of Sciences of the Czech Republic. Hinode is a Japanese mission developed and launched by ISAS/JAXA, collaborating with NAOJ as a domestic partner, NASA and STFC (UK) as international partners. Scientific operation of the Hinode mission is conducted by the Hinode science team organized at ISAS/JAXA. This team mainly consists of scientists from institutes in the partner countries. Support for the post-launch operation is provided by JAXA and NAOJ (Japan), STFC (U.K.), NASA, ESA, and NSC (Norway). This research has made use of NASA's Astrophysics Data System. We thank an anonymous referee for suggestions that lead to important improvements on the readability of the manuscript.
\end{acknowledgements}

\bibliographystyle{aa}
\bibliography{aa54241-25}

\FloatBarrier

\begin{appendix}

\section{Do MHS inversions enforce \texorpdfstring{$\nabla P_{\rm g} = \ve{L}$}{∇ Pg = L} ?}
\label{sec:mhs}

Since the MHS inversions carried out with the FIRTEZ Stokes inversion code take into account the Lorentz force, it is natural to wonder whether our observational results verify $r^{-1} \partial P_{\rm g} / \partial \phi = L_\phi$ as a consequence of being internally imposed. In order to answer this question, it is convenient to take a step back and first study the implications of having hydrostatic equilibrium (HE). Full (i.e. in 3D) hydrostatic equilibrium is obtained by neglecting the Lorentz force ($\ve{L}=0$ in Eq.~\ref{eq:mhs_all}). The resulting equation can be decomposed in:

\begin{eqnarray} 
\frac{\partial P_{\rm g}}{\partial x} = \frac{\partial P_{\rm g}}{\partial y} = 0\label{eq:hexy} \\
\frac{\partial P_{\rm g}}{\partial z} = -\rho g\label{eq:hez}
\end{eqnarray}

With the exception of the FIRTEZ code, all Stokes inversion codes such as SIR \citep{basilio1992sir}, SPINOR \citep{frutiger1999spinor}, STiC \citep{jaime2019stic}, NICOLE \citep{socas2015nicole}, etc. consider only Eq.~\ref{eq:hez} (i.e. vertical hydrostatic equilibrium), but they neglect that there is also horizontal hydrostatic equilibrium (Eqs.~\ref{eq:hexy}) that implies that the gas pressure cannot 
vary in planes of constant $z$:

\begin{equation}
P_{\rm g} \ne f(x,y)
\end{equation}

Consequently, two pixels on the solar surface, sitting next to each other (on the XY plane) must have the same $P_{\rm g}(z)$. This must be satisfied $\forall z$'s. Next, by applying Eq.~\ref{eq:hez}, we can also conclude that all pixels on the XY plane must have the same density $\rho(z)$, and therefore also the same temperature $T(z)$. In other words, neither $\rho$, $P_{\rm g}$ nor $T$ can depend on $(x,y)$. Now, we must take into account that typically from the inversion of the Stokes vector we obtain $T(x,y,z)$. This immediately implies that horizontal hydrostatic equilibrium cannot be maintained in the sense that Eqs.~\ref{eq:hexy} and ~\ref{eq:hez} cannot be simultaneously satisfied. Hence Eq.~\ref{eq:hexy} is usually ignored and only Eq.~\ref{eq:hez} is considered, hence the term vertical hydrostatic equilibrium.\\

Now, as explained in \citet{borrero2019firtez}, FIRTEZ does not directly solve the momentum equation in magneto-hydrostatics (Eq.~\ref{eq:mhs_all}). Instead, it takes the divergence of that equation and iteratively solves the following second-order Poisson equation:

\begin{equation}
\nabla^2 (\ln P_{\rm g}) = - \frac{m_a g}{K_b} \frac{\partial}{\partial z}\left[\frac{u}{T}\right] + \nabla \cdot \left(\frac{\ve{L}^{\prime}}{P_{\rm g}}\right) \label{eq:poisson_mhs}
\end{equation}

\noindent where $\ve{L}^{\prime}$ is a modified Lorentz force. Let us leave $\ve{L}^{\prime}$ aside for the moment and ask ourselves what does FIRTEZ do when the Lorentz force is neglected (i.e. hydrostatic equilibrium). Under this assumption FIRTEZ solves:

\begin{equation}
\nabla^2 (\ln P_{\rm g}) = - \frac{m_a g}{K_b} \frac{\partial}{\partial z}\left[\frac{u}{T}\right] \label{eq:poisson_heq}
\end{equation}

But let us now remember that typically the inversion will yield $T(x,y,z)$ and hence also a variable mean molecular weight $u(x,y,z)$, so the that the term:

\begin{equation}
\frac{\partial}{\partial z}\left[\frac{u}{T}\right] = f(x,y,z)
\end{equation}

And consequently, the resulting $P_{\rm g}$ from solving Eq.~\ref{eq:poisson_heq} will also be a function of $(x,y,z)$. But this is contradictory with hydrostatic equilibrium as we just saw above because full 3D hydrostatic equilibrium implies $P_{\rm g}=f(z)$ only.\\

\begin{figure*}[ht!]
\begin{center}
\begin{tabular}{ccc}
\includegraphics[height=4cm]{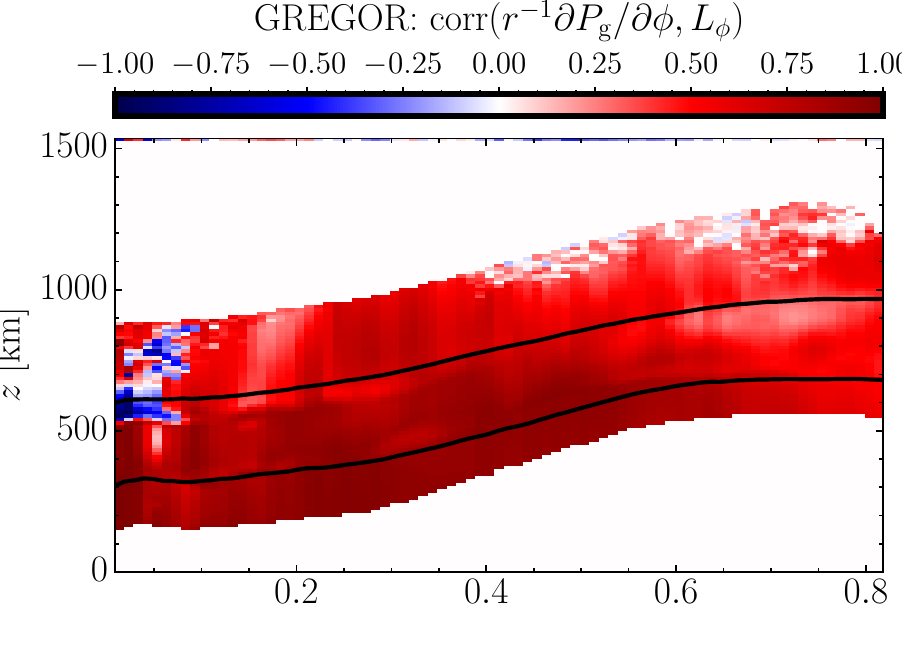} &
\includegraphics[height=4cm]{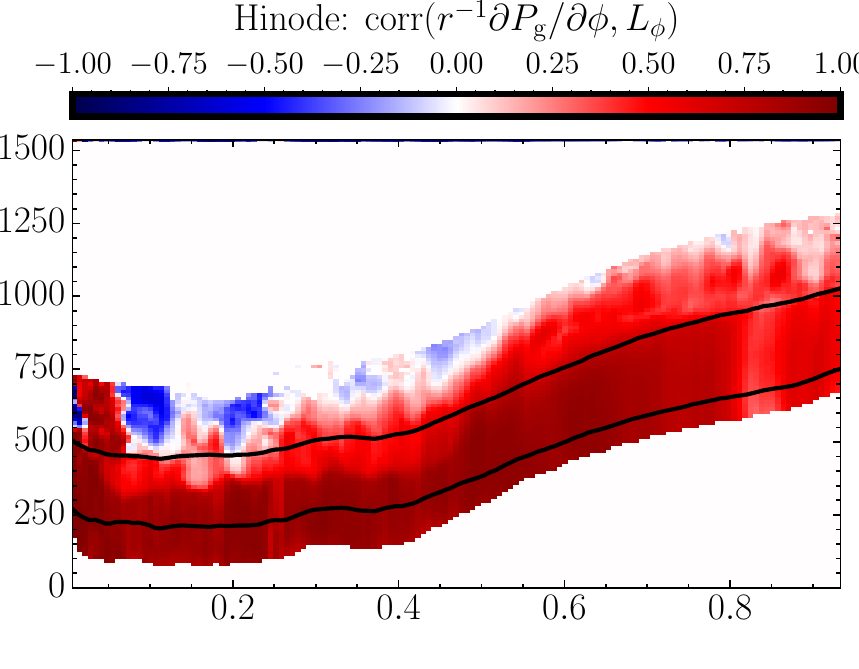} &
\includegraphics[height=4cm]{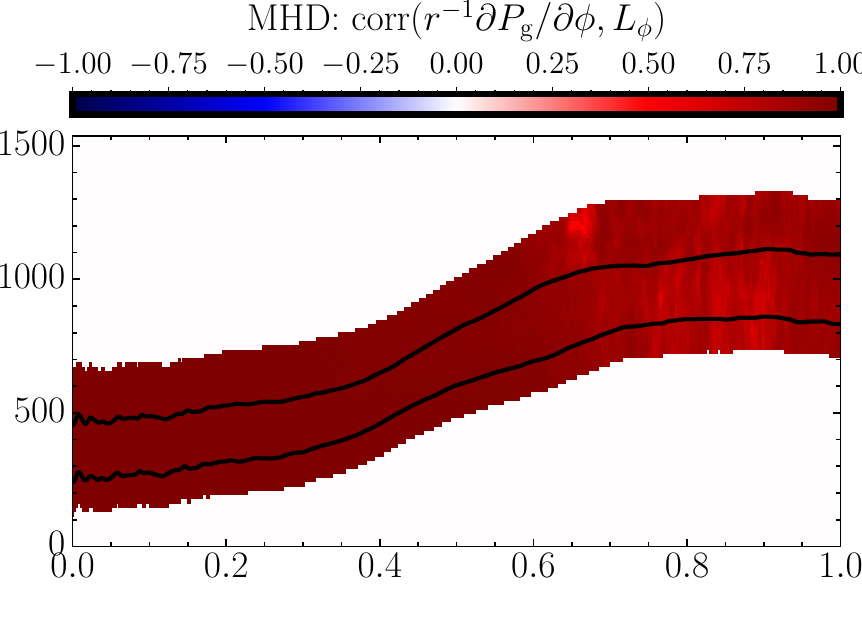} \\
\includegraphics[height=4cm]{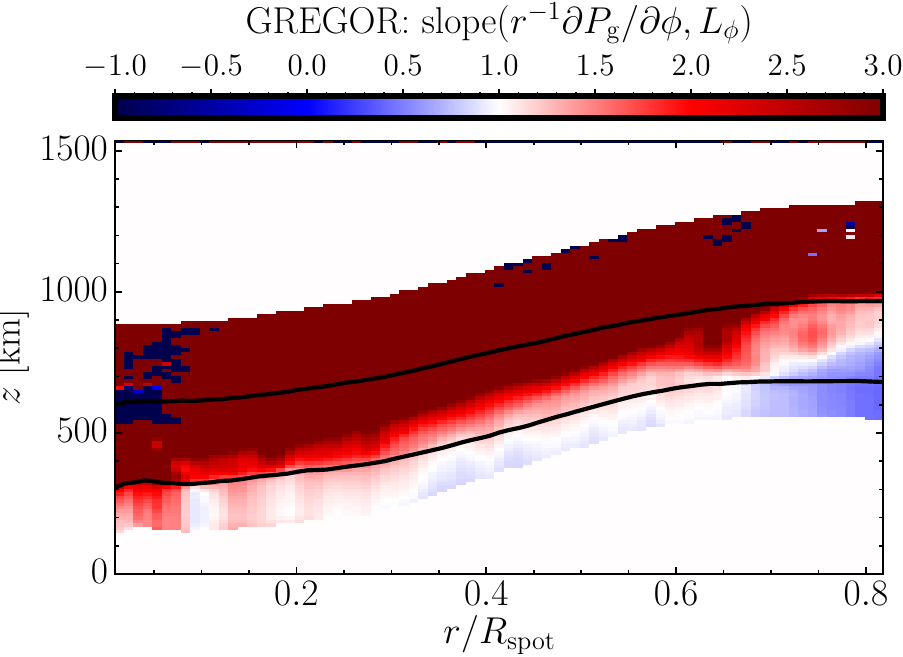} &
\includegraphics[height=4cm]{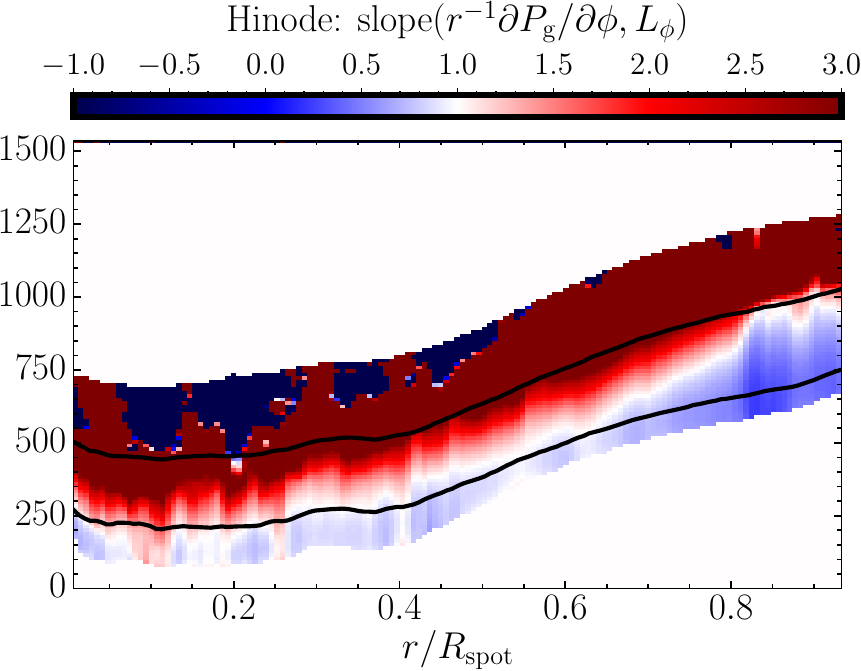} &
\includegraphics[height=4cm]{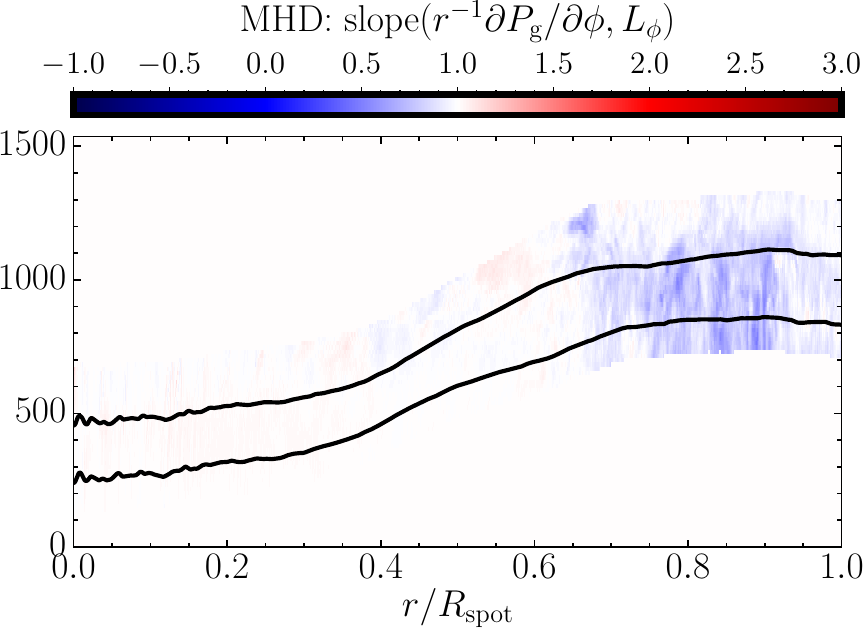} \\
\end{tabular}
\caption{Study of the MHS equilibrium along the azimuthal direction for the whole sunspots. Top panels: Pearson's correlation coefficient between the rhs and lhs of Eq.~\ref{eq:eq_phi} as a function of $(r,z)$. Bottom panels: same as on the top but for the slope of the linear regression between the rhs and lhs of Eq.~\ref{eq:eq_phi} as a function of $(r,z)$. $R_s$ represents the radius of the sunspots. Solid black lines show the $\phi$-average location of the $\tau_{\rm c}=1,10^{-2}$. Observational results are displayed on the left (GREGOR) and middle (Hinode) panels. Results from the MHD simulations are presented in the rightmost panels.\label{fig:corr_rz}}
\end{center}
\end{figure*}

So, it is clear that the method FIRTEZ employs to retrieve the gas pressure will result in $P_{\rm g} = f(x,y,z)$ even if $\ve{L}^{\prime} = 0$ and thus, the fact that $P_{\rm g}$ changes horizontally (i.e. along $\phi$), is not necessarily due to the Lorentz force. In fact, there is no guarantee that the following equation:

\begin{equation}
L_\phi=r^{-1}\frac{\partial P_{\rm g}}{\partial \phi}
\label{eq:eq_phi}
\end{equation}

\noindent must be satisfied always at all points of the three-dimensional domain. In order to prove this particular point, the top panels on Figure~\ref{fig:corr_rz} show the correlations between the right-hand side and left-hand side of Eq.~\ref{eq:eq_phi} as a function of $(r,z)$ in the observed and simulated sunspots.\\

Figure~\ref{fig:corr_rz} shows that although the correlations are large in the observations (left and middle panels for GREGOR and Hinode, respectively), this is certainly not always the case. In fact there are regions where the obtained correlation is actually negative. Such is the case of the high photosphere at all radial distances, and of the mid-photosphere at short radial distances (i.e. in the umbra). For comparison the simulated sunspot is also displayed in the rightmost panel. Here we can see that the MHD simulations show a high degree of correlation over a wider range of radial distances $r/R_{\rm spot}$ and heights $z$.\\

On the other hand, a high correlation between the rhs and lhs side of Eq.~\ref{eq:eq_phi} does not necessarily mean that this equation is strictly fulfilled. It can occur that they are correlated but not necessarily equal. We have evaluated this by performing a linear fit to the scatter plot of $L_\phi$ and $r^{-1}\partial P_{\rm g}/\partial \phi$ (as done in panel d of Fig.~\ref{fig:phi}) and plotting on the bottom panels of Figure~\ref{fig:corr_rz}, the slope $m$ of the linear fit between the rhs and the lhs of Eq.~\ref{eq:eq_phi} as a function of $(r,z)$.\\

Figure~\ref{fig:corr_rz} (bottom-right and bottom-middle panels) shows that even though there might be very high correlations between the two terms on either side of Eq.~\ref{eq:eq_phi} the actual equation is strictly fulfilled only in the region where the slope is close to 1.0, that is, in a narrow region of the observed sunspots around the $\tau_c=1$ level. This is in contrast with the sunspot simulated with the MuRAM code, where Eq.~\ref{eq:eq_phi} is verified to a large degree (i.e. slope 1.0) over a much wider region, both vertically and radially, than in the observations analyzed with the FIRTEZ code. Because of these results it is possible to state that Eq.~\ref{eq:eq_phi} is not strictly imposed by FIRTEZ.\\

The question now arises to what extent we can expect Eq.~\ref{eq:eq_phi} to be strictly fulfilled. As explained above, $T(x,y,z)$ implies that full 3D hydrostatic equilibrium cannot be fulfilled by FIRTEZ even if the Lorentz force is zero. This can be also interpreted in the following way: since under hydrostatic equilibrium (i.e. $\ve{L}^{\prime}=0$) $T$ and $P_{\rm g}$ can only a function of $z$, the fact that FIRTEZ receives $T(x,y,z)$ already implies the presence of a residual force not given by $(\nabla \times \ve{B})\times \ve{B}$ (i.e. the Lorentz force) so that the resulting gas pressure also depends on $(x,y,z)$.\\

The nature of the aforementioned residual force can be understood by looking at  the right-hand-side of Eq.~\ref{eq:poisson_mhs}. Here, after defining $\ve{F} = \ve{L}^{\prime}/P_{\rm g}$ and projecting onto its Hodge's components: $\ve{F} = \nabla \Psi + \nabla \times \ve{W}$, we see that Eq.~\ref{eq:poisson_mhs} becomes:

\begin{equation}
\nabla^2 (\ln P_{\rm g}) = - \frac{m_a g}{K_b} \frac{\partial}{\partial z}\left[\frac{u}{T}\right] + \nabla^2 \Psi \label{eq:poisson_hodge}
\end{equation}

\noindent where it is clear now that in the presence of a magnetic field, FIRTEZ infers a gas pressure $P_{\rm g}$ that balances the potential part $\Psi$ of the force $\ve{F} = \ve{L}^{\prime}/P_{\rm g}$ with the residual forces being identified with the non-potential part $\ve{W}$ of this force.\\

Therefore, one can expect the results from the FIRTEZ code to fulfill Eq.~\ref{eq:eq_phi} when the potential part of the $\ve{L}^{\prime}/P_{\rm g}$ actually dominates over its non-potential component. Otherwise it is conceivable that the potential component will still be capable of introducing a correlation between $L_\phi$ and $r^{-1} \partial P_{\rm g}/\partial \phi$ but the slope of the linear fit between these two quantities might be quite different from 1.0 (see Fig.~\ref{fig:corr_rz}).\\

It is now time to remember that the modified Lorentz force $\ve{L}^{\prime}$ used by FIRTEZ to determine the gas pressure via the solution of Eq.~\ref{eq:poisson_mhs} is not necessarily the Lorentz force obtained via the inferred magnetic field. Instead, FIRTEZ takes the calculated electric current $\ve{j} = c (4\pi)^{-1} \nabla \times \ve{B}$ and dampens it by a factor $\propto \beta^{2}$ (with $\beta$ being the ratio between magnetic and gas pressure) whenever the module $\|\ve{j}\|$ exceeds a maximum allowed value. This is done with the aim that, in the upper atmosphere where the $\beta \ll 1$ and where spectral lines do not convey accurate values of magnetic field $\ve{B}$, electric currents determined from these inaccurate magnetic fields do not dominate the force balance. This explains why the agreement of the observational results with Eq.~\ref{eq:eq_phi} worsens progressively from the lower atmosphere towards the upper atmosphere (see Fig.~\ref{fig:corr_rz}). This effect reinforces the previous conclusion that the MHS equilibrium imposed by FIRTEZ does not strictly impose Eq.~\ref{eq:eq_phi}.\\

\section{Simulation details}
\label{sec:app_sim}

The initial magnetic field of the simulation uses the self-similar approach used in \cite{1958IAUS....6..263S} and others:

\begin{align}
    B_z(r,z)&=B_0f(\xi)g(z),\\
    B_r(r,z)&=-B_0\frac{r}{2}f(\xi)g'(z),\\
    \xi&=r\sqrt{g(z)},
\end{align}

whereby $B_z$ is the vertical component of the magnetic field, $B_r$ the radial component, $r$ the distance from the spot axis, and $z$ the distance from the bottom of the box. For this simulation

\begin{align}
    f(\xi)&=\exp\left(-(\xi/R_0)^4\right),\\
    g(z)&=\exp\left(-(z/z_0)^2\right).
\end{align}

 The parameters are $B_0=6.4\,$kG, the initial field strength on the spot axis at the bottom of the box, $R_0=8.2\,$Mm and $z_0=6.4185\,$Mm, which determines how far away from the axis and the bottom $B_\mathrm{z}$ reduces by a factor $e$. These parameters lead to a sunspot with an initial flux of $F_0=1.2\times10^{22}$Mx and a field strength at the top of 2.56\,kG. A more detailed description of the initial state of the simulation is in \cite{rempel2012mhd}. Other self-similar sunspot models use for $f(\xi)$ a Gaussian \citep[e.g.][]{1958IAUS....6..263S,2005A&A...441..337S} or the boxcar function \citep[many analytical models and e.g.][]{2020ApJ...893..113P} and $g(z)$ is sometimes related to the pressure difference between the spot center and the quiet Sun \citep[e.g.][]{1958IAUS....6..263S}.\\

The parameter $\alpha=2$ means that at the top of the box, the horizontal field is twice what it would be if potential extrapolation were employed:

\begin{align}
B_z & = \hphantom{\alpha\ }\mathcal{F}^{-1} \left(\mathcal{F}(B_{z,0})\mathrm{e}^{-z\,\alpha\,|k|}\right)
\label{eq:bz_alpha}\\
\intertext{and}
B_x & = \alpha \mathcal{F}^{-1} \left(\mathcal{F}(B_{z,0}) \frac{-ik_x}{|k|} \mathrm{e}^{-z\,\alpha\,|k|}\right),
\label{eq:bx_alpha}
\end{align}

whereby $B_{z,0}$ is the vertical field at the top of the box, $z$ the height above the top of the box (in the ghost cells), $\mathcal{F}$ the horizontal Fourier transform, $k_x$ the Fourier modes, and $|k|=\sqrt{\smash[b]{k_x^2+k_y^2}}\vphantom{k_x}$. For $B_y$ holds the same equation as Eq.~\ref{eq:bx_alpha} but performing the substitution $k_x \rightarrow k_y$. $\alpha>1$ is responsible for $B_r$ being larger than in the simulated penumbra compared to observations.\\

For a comparison to observed sunspots, see \citet{jan2020sunspot}, which concludes that in sunspots with $\alpha>1$ have a too low $B_\mathrm{z}$ at the umbral boundary. Furthermore, \citet{2021A&A...656A..92S} conclude, that a sufficiently strong $B_\mathrm{z}$ suppresses penumbral-type overturning convection.\\

\section{Additional results}
\label{sec:add}

\begin{figure*}[ht!]
\begin{center}
\begin{tabular}{cc}
\includegraphics[width=7cm]{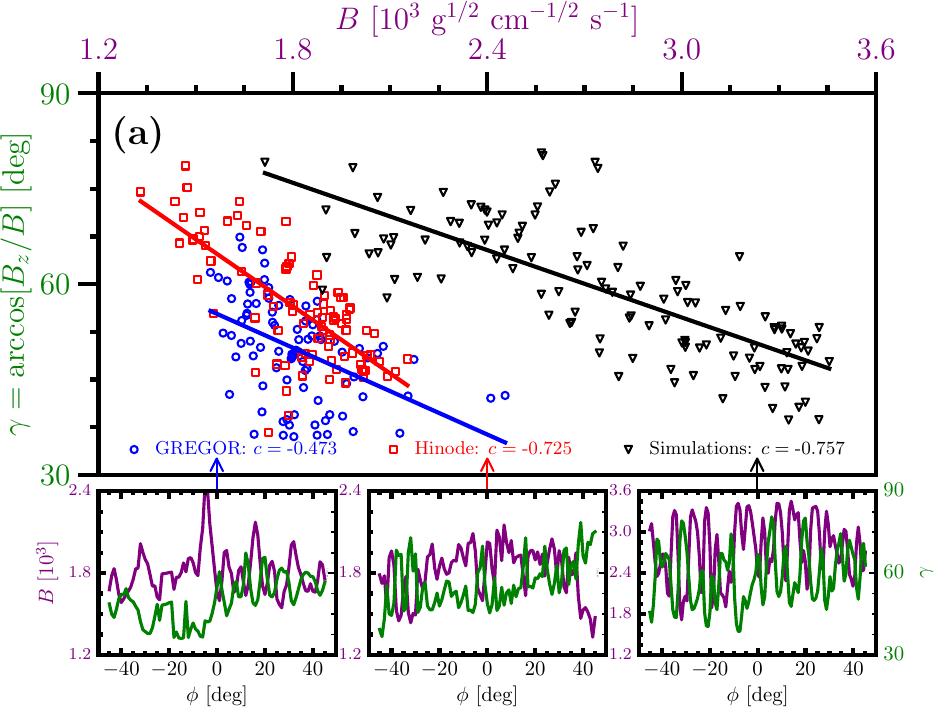} &
\includegraphics[width=7cm]{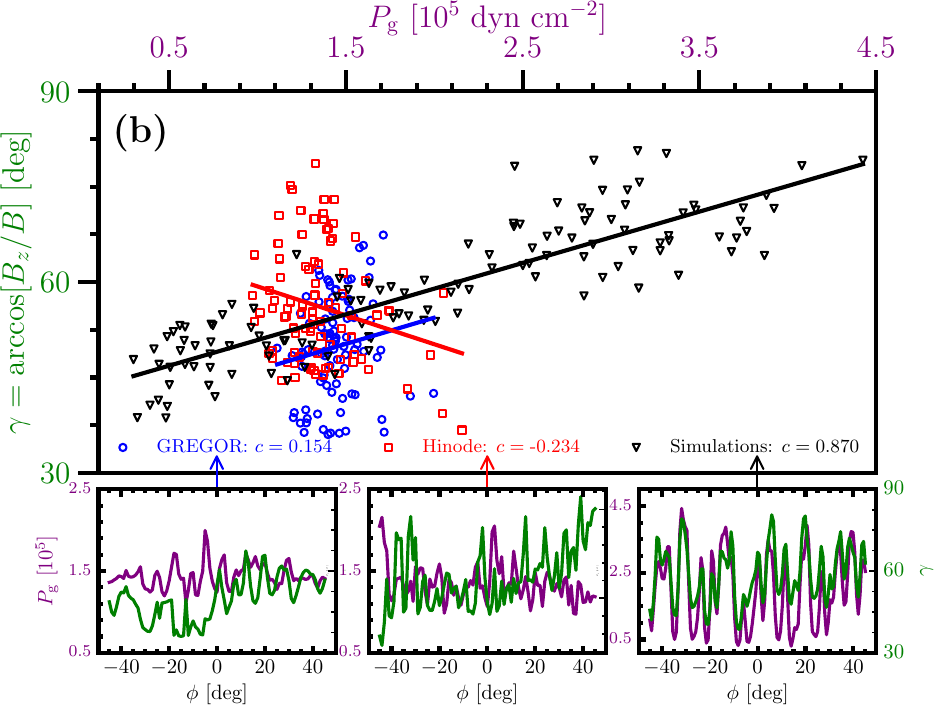} \\
\includegraphics[width=7cm]{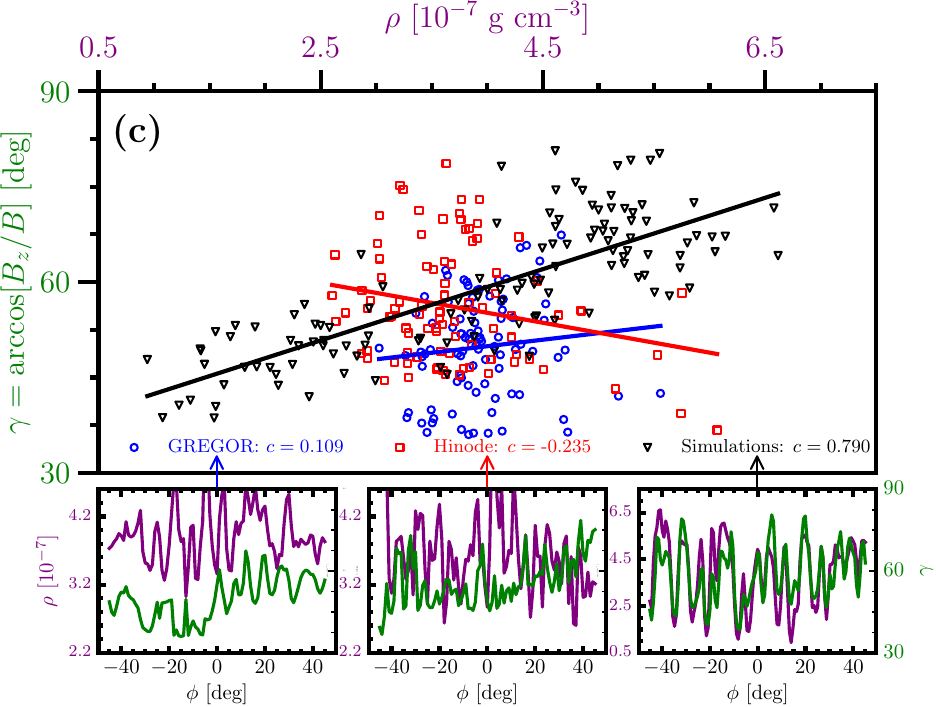} &
\includegraphics[width=7cm]{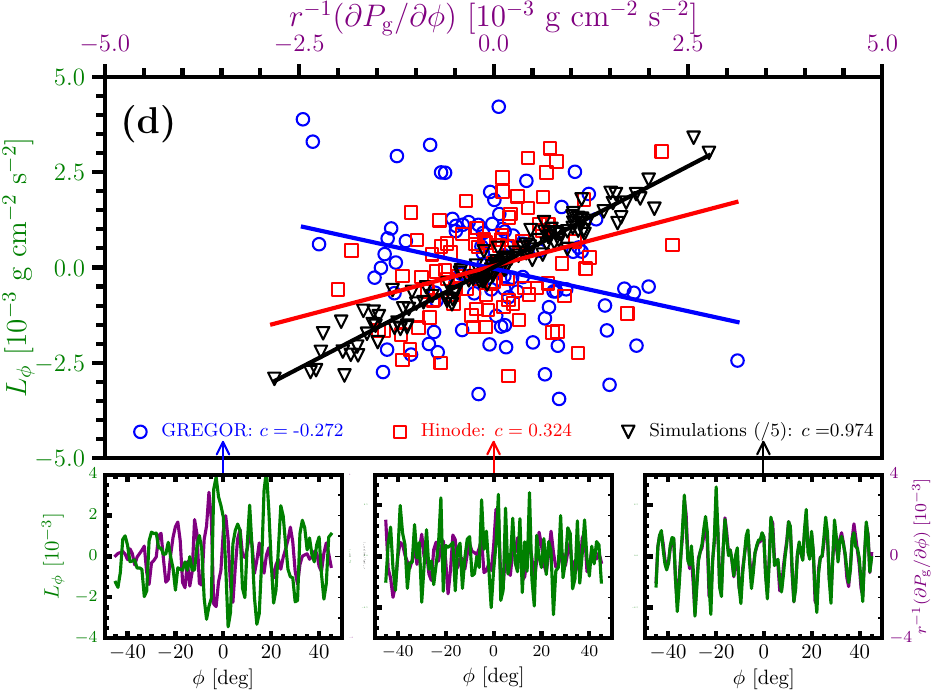} 
\end{tabular}
\caption{Same as Figure~\ref{fig:phi} but obtained under the assumption of vertical hydrostatic equilibrium (see Sect.~\ref{sec:hydro}). Results from MHD simulations (black colors and right-most minipanels) are identical to Fig.~\ref{fig:phi}.\label{fig:phi_hydro}}
\end{center}
\end{figure*}

\begin{figure*}[ht!]
\begin{center}
\begin{tabular}{cc}
\includegraphics[width=7cm]{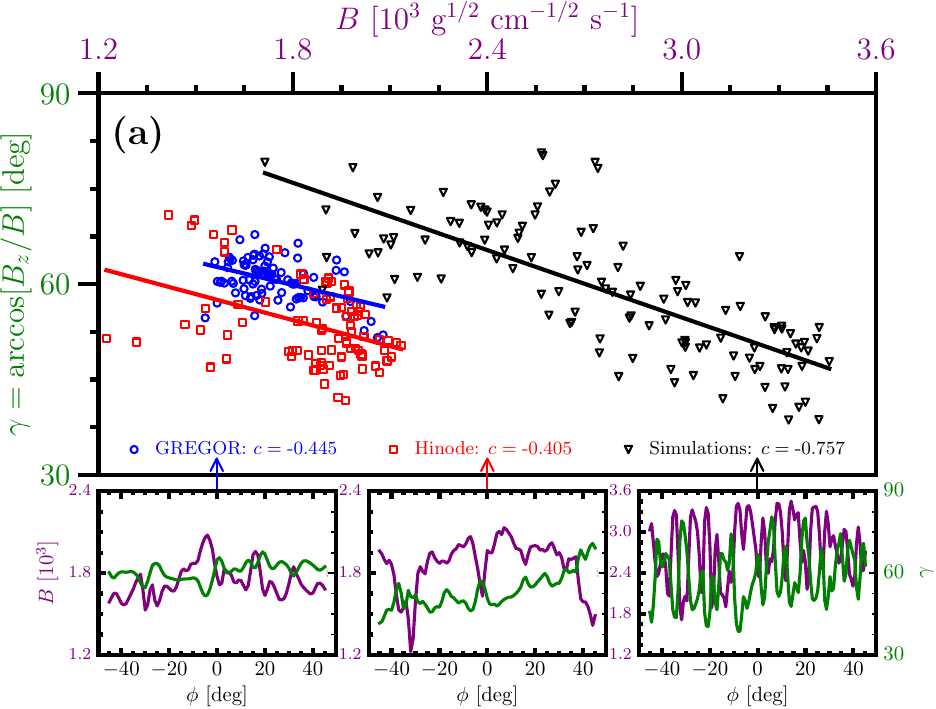} &
\includegraphics[width=7cm]{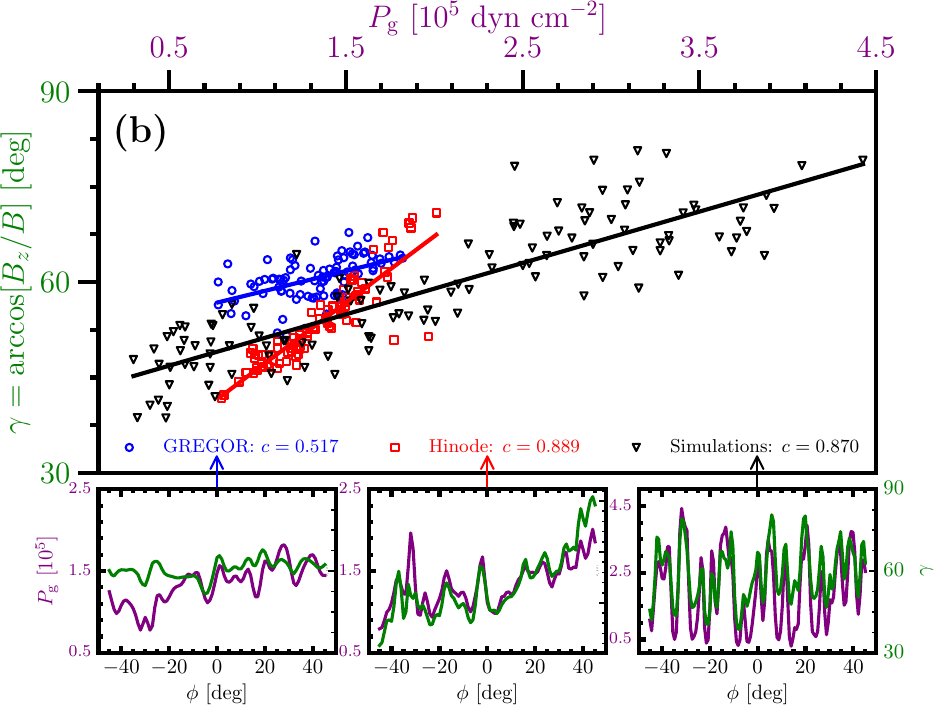} \\
\includegraphics[width=7cm]{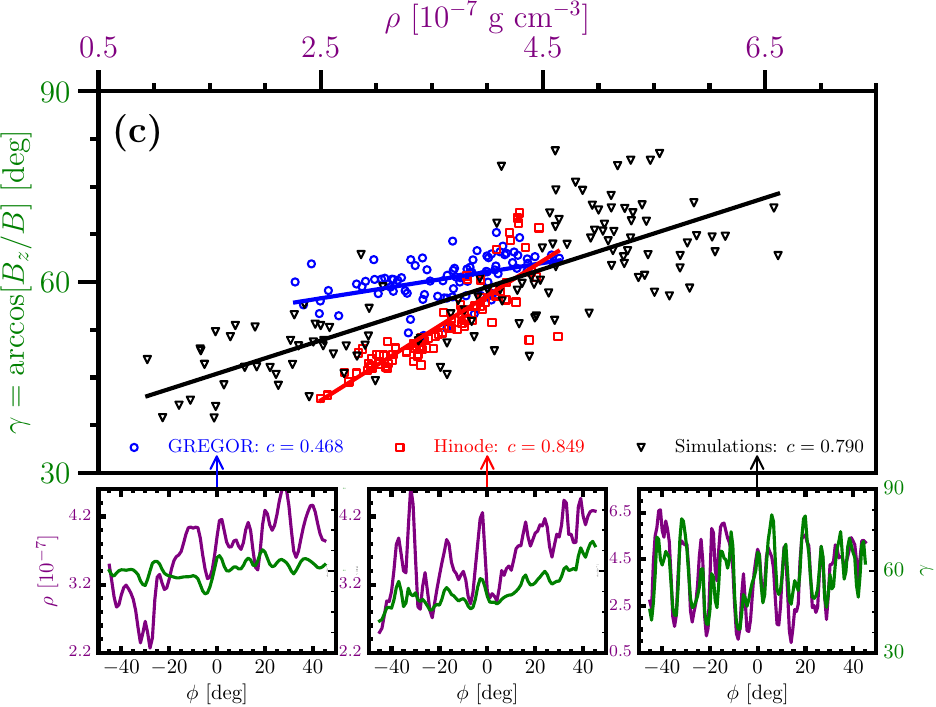} &
\includegraphics[width=7cm]{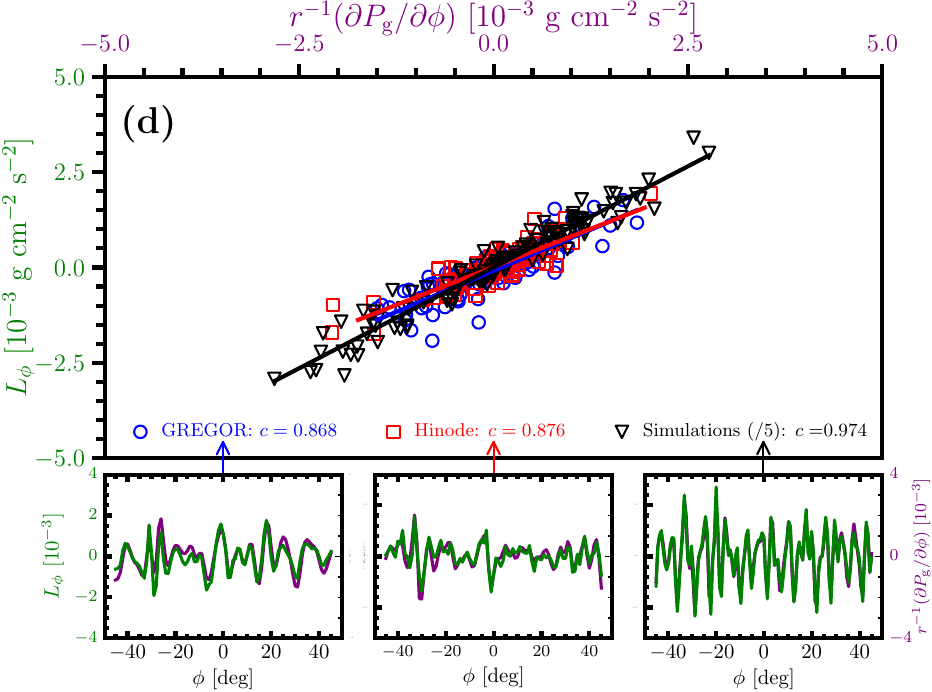} 
\end{tabular}
\caption{Same as Figure~\ref{fig:phi} but obtained without horizontal coupling between the pixels due to the telescope's PSF (see Sect.~\ref{sec:res_uncoupled}). Results from MHD simulations (black colors and right-most minipanels) are identical to Fig.~\ref{fig:phi}\label{fig:phi_pixelwise}}
\end{center}
\end{figure*}

\end{appendix}
\end{document}